\documentclass[sigconf,natbib=false]{acmart}
\pdfoutput=1

\usepackage[
    backend=biber,
    datamodel=acmdatamodel,
    bibstyle=acmnumeric,
    citestyle=numeric-comp,
    urldate=comp,
]{biblatex}
\usepackage{xspace}
\usepackage{amsthm}
\usepackage{multirow}
\usepackage{tabularx}
\usepackage{multirow}
\usepackage{makecell}
\usepackage{multicol}
\PassOptionsToPackage{hyphens}{url}
\usepackage{float}
\usepackage{adjustbox}
\usepackage{xstring}
\usepackage{soul}
\usepackage[shortlabels,inline]{enumitem}
\usepackage{cleveref}
\usepackage[frozencache,cachedir=.]{minted}
\usepackage[english=american,strict]{csquotes}
\MakeOuterQuote{"}
\usepackage[
    group-separator={,},
    group-minimum-digits=4,
]{siunitx}
\usepackage{balance}

\usepackage{fourier}
\usepackage{colortbl}
\usepackage{titlesec}
\usepackage{threeparttable}

\usepackage{times}

\newcommand{\codebookquote}[3][]{#3}

\copyrightyear{2024}
\acmYear{2024}

\acmConference[CCS '24]{ACM Conference on Computer and Communications Security}{October 14-18, 2024}{Salt Lake City, U.S.A.}

\usepackage{listings}

\definecolor{lstgreen}{rgb}{0,0.6,0}

\lstdefinestyle{plain}{%
  numbers=none,
  frame=none,
  xleftmargin=1pt,
  xrightmargin=1pt,
}

\lstdefinelanguage{Rust}{%
  morekeywords={as,break,const,continue,crate,else,enum,extern,false,
    fn,for,if,impl,in,let,loop,match,mod,move,mut,pub,ref,return,Self,
    self,static,struct,super,trait,true,type,unsafe,use,where,while,
    abstract,alignof,become,box,do,final,macro,offsetof,override,priv,
    proc,pure,sizeof,typeof,unsized,virtual,yield},
  morekeywords=[2]{isize,usize,char,bool,str,String,u8,u16,u32,u64,u128,i8,i16,i32,i64,i128,f32,f64},
  sensitive=true,
  morecomment=[l]{//},
  morecomment=[l]{///}, 
  morestring=[b]{"},
}%

\setminted{
  escapeinside=||,
  fontsize=\scriptsize,
  frame=single,
  numbers=left,
  numbersep=1ex,
}

\usepackage{tikz}
\usetikzlibrary{shapes,arrows,positioning,shapes.multipart}

\let\underscore\_
\renewcommand{\_}{$\underscore$\allowbreak{}}

\newlength{\nowidthtmp}

\renewcommand{\paragraph}[1]{\par\medskip\noindent\textbf{#1.}\hspace{1ex minus .1ex}}
\renewcommand{\subparagraph}[1]{\par\smallskip\noindent\textit{#1.}\hspace{.5ex minus .1ex}}

\let\oldsubsubsection\subsubsection
\renewcommand{\subsubsection}[1]{\oldsubsubsection{#1}\leavevmode\par\noindent\ignorespaces}

\newcommand{\citewauthor}[1]{\citeauthor*{#1}~\cite{#1}}

\ExplSyntaxOn
\NewDocumentCommand\pnum{mm}{
    \fp_set:Nn\l_a{#1}
    \fp_set:Nn\l_b{#2}

    \num{\fp_eval:n{\l_a}}\nobreak{}~(\SI{\fp_eval:n{round(\l_a/\l_b*100, 1)}}{\percent})
}

\NewDocumentCommand\p{O{100} O{0} m}{
    \fp_set:Nn\l_base{#1}
    \fp_set:Nn\l_prec{#2}
    \fp_set:Nn\l_val{#3}

    \int_compare:nNnTF {#1} = {100} {
        \int_compare:nNnTF {#2} = {0} {
            \SI{#3}{\percent}
        }{
            \SI{\fp_eval:n{round(\l_val/\l_base*100, \l_prec)}}{\percent}
        }
    }{
        \SI{\fp_eval:n{round(\l_val/\l_base*100, \l_prec)}}{\percent}
    }
}
\ExplSyntaxOff

\usepackage{tikz}
\usepackage[tikz,usetwoside=true]{mdframed}
\usetikzlibrary{calc,arrows}
\tikzset{
    summary arrow/.style={%
        line width=1pt,
        draw=gray!40,
        rounded corners=1ex,
    },
    summary head/.style={
        fill=white,
        font=\bfseries\sffamily,
        text=gray!80,
        anchor=base west,
    },
}

\mdfdefinestyle{summary}{%
    singleextra={%
        \path let \p1=(P), \p2=(O) in (\x2,\y1) coordinate (Q);
        \path let \p1=(P), \p2=(O) in (\x1,\y2) coordinate (R);
        \path let \p1=(Q), \p2=(O) in (\x1,{(\y1-\y2)/2}) coordinate (M);
        \path [summary arrow] (R) -| (M) |- (P);
        \node [summary head] at ($(Q)+(1ex,-1.5pt)$) {\frametitle};
    },
    firstextra={%
        \path let \p1=(P), \p2=(O) in (\x2,\y1) coordinate (Q);
        \path [summary arrow,latex-] (O) |- (P);
        \node [summary head] at ($(Q)+(1ex,-1.5pt)$) {\frametitle};
    },
    secondextra={%
        \path let \p1=(P), \p2=(O) in (\x2,\y1) coordinate (Q);
        \path let \p1=(P), \p2=(O) in (\x1,\y2) coordinate (R);
        \path [summary arrow] (R) -| (Q);
    },
    middleextra={%
        \path let \p1=(P), \p2=(O) in (\x2,\y1) coordinate (Q);
        \path [summary arrow] (O) -- (Q);
    },
    middlelinewidth=1.5em,middlelinecolor=white,
    hidealllines=true,topline=true,
    innertopmargin=0pt,
    innerbottommargin=1em,
    innerrightmargin=0pt,
    innerleftmargin=1.5ex,
    skipabove=0.5\baselineskip,
    skipbelow=0.3\baselineskip,
}

\newenvironment{summary}[1]{
\begingroup
\def\frametitle{\small #1}
\begin{mdframed}[style=summary]
}{
\end{mdframed}
\endgroup
}

\DeclareQuoteStyle[american]{english}
    {\itshape\textquotedblleft}
    {\textquotedblright}
    [0.05em]
    {\textquoteleft}
    {\textquoteright}

\newcommand{\participantquote}[3][]{\blockquote[P{#2}]{#3}} 
\newcommand{\surveyquote}[3][]{\blockquote[\Ps{#2}]{#3}}
\newcommand{\simplequote}[1]{\enquote{#1}}

\newcommand{\rnum}[1]{({#1})\nolinebreak{}}
\newcommand{\Ps}[1]{P\textsubscript{s}#1}

\newcounter{causescounter}
\renewcommand*{\thecausescounter}{C\arabic{causescounter}}
\newcommand{\cause}[1]{\refstepcounter{causescounter}\paragraph{\thecausescounter: #1}}
\Crefname{causescounter}{}{}

\newcommand{\ddd}{\dots~}

\newcommand{\ssc}{Solana smart contracts\xspace}

\newcommand{\iobug}{{IB}\xspace}
\newcommand{\acpi}{{ACPI}\xspace}
\newcommand{\msc}{{MSC}\xspace}

\newcommand{\symresDataSize}{\num{6324}\xspace}
\newcommand{\symresACPI}{94\xspace}
\newcommand{\symresOwner}{78\xspace}
\newcommand{\symresSigner}{24\xspace}
\newcommand{\symresVuln}{14\xspace}

\usepackage{hyperref}
\hypersetup{
    colorlinks,
    linkcolor={black},
    citecolor={red!70!black},
    urlcolor={blue!70!black}
}

\begin{document}

\title{Defying the Odds: Solana's Unexpected Resilience in Spite of the Security Challenges Faced by Developers}

\author{Sébastien Andreina}
\email{sebastien.andreina@neclab.eu}
\affiliation{%
  \institution{NEC Laboratories Europe}
  \city{Heidelberg}
  \country{Germany}
}

\author{Tobias Cloosters}
\email{tobias.cloosters@uni-due.de}
\affiliation{%
  \institution{University of Duisburg-Essen}
  \city{Essen}
  \country{Germany}}

\author{Lucas Davi}
\email{lucas.davi@uni-due.de}
\affiliation{%
  \institution{University of Duisburg-Essen}
  \city{Essen}
  \country{Germany}}

\author{Jens-Rene Giesen}
\email{jens-rene.giesen@uni-due.de}
\affiliation{%
  \institution{University of Duisburg-Essen}
  \city{Eseen}
  \country{Germany}}

\author{Marco Gutfleisch}
\email{marco.gutfleisch@rub.de}
\affiliation{%
  \institution{Ruhr University Bochum}
  \city{Bochum}
  \country{Germany}}

\author{Ghassan Karame}
\email{ghassan@karame.org}
\affiliation{%
  \institution{Ruhr University Bochum}
  \city{Bochum}
  \country{Germany}}

\author{Alena Naiakshina}
\email{alena.naiakshina@rub.de}
\affiliation{%
  \institution{Ruhr University Bochum}
  \city{Bochum}
  \country{Germany}}

\author{Houda Naji}
\email{houda.naji@rub.de}
\affiliation{%
  \institution{Ruhr University Bochum}
  \city{Bochum}
  \country{Germany}}
\renewcommand{\shortauthors}{Andreina, Et Al.}

\date{}

\ccsdesc[500]{Security and privacy~Distributed systems security}
\ccsdesc[500]{Software and its engineering~Software safety}


\begin{abstract}
Solana gained considerable attention as one of the most popular blockchain platforms for deploying decentralized applications. Compared to Ethereum, however, we observe a lack of research on how Solana smart contract developers handle security, what challenges they encounter, and how this affects the overall security of the ecosystem.
To address this, we conducted the first comprehensive study on the Solana platform {consisting of a 90-minute Solana smart contract code review task with 35 participants followed by interviews with a subset of seven participants.} Our study shows, quite alarmingly, that none of the participants could detect all important security vulnerabilities in a code review task and that \p{83} of the participants are likely to release vulnerable smart contracts.
Our study also sheds light on the root causes of developers' challenges with Solana smart contract development, suggesting the need for better security guidance and resources. In spite of these challenges, our automated analysis on currently deployed Solana smart contracts surprisingly suggests that the prevalence of vulnerabilities---especially those pointed out as the most challenging in our developer study---is below 0.3\%. We explore the causes of this counter-intuitive resilience and show that frameworks, such as Anchor, are aiding Solana developers in deploying secure contracts.
\end{abstract}

\maketitle


\section{Introduction}
With the launch of the Solana blockchain in 2020, Solana has quickly grown to become one of the most widely used platforms for deploying decentralized applications (DApps) and non-fungible tokens (NFTs). 
At the time of writing, Solana is among the top 10 blockchain platforms in terms of total market capitalization~\cite{coinmarketcap}. 
Solana's popularity has been further heightened by its low fees compared to its main competitors, such as Ethereum and Polygon. Solana transaction fees are 2 to 3 orders of magnitude cheaper than those incurred in Ethereum~\cite{solana-fees}. 
Unlike Ethereum's monolithic smart contract execution environment (the EVM), Solana decouples the execution logic from the state of the smart contract and relies on the programming language Rust.
Although Rust poses certain challenges for developers~\cite{fulton2021benefits,zhu2022learning, crichton2020usability}, its security features have been praised and increased developers' confidence during development~\cite{fulton2021benefits}. 

With the Solana execution environment, new vulnerability patterns were introduced that are not captured by existing EVM-based analysis tools~\cite{neodyme-common-pitfalls}.
One of the most recent major hacks, the Wormhole hack~\cite{wormhole}, resulted in the loss of 325~million~USD due to a missing key check issue.
In particular, attacks exploiting missing ownership and signer checks (that exploit the lack of validation to bypass access control), and cross-program invocation (that exploits the lack of call target validation during cross account calls), seem to be quite intrinsic to the very nature of the Solana execution environment~\cite{neodyme-common-pitfalls}. 

Compared to Ethereum, there is a lack of understanding of why these vulnerabilities exist, how Solana smart contract developers handle security, what challenges they encounter, and how this affects the overall security of the ecosystem. 

In this work, we set forth to understand the challenges faced by developers during the development of Solana smart contract and to explore the following research questions:

\begin{enumerate}[
    label={RQ\arabic*:},
    format=\textbf,
    wide,
    labelindent=0pt,
    topsep=2pt,
    leftmargin=1em,
    noitemsep,
]

\item \textit{Do Solana smart contract developers recognize prominent security vulnerabilities in smart contracts?} 

\item \textit{What challenges do developers encounter that impact the development of secure smart contracts?}

\item \textit{Given these challenges, what is the prevalence of vulnerabilities in Solana smart contracts?}

\end{enumerate}

\paragraph{Comprehensive Study on Developer Security Practices}
To address RQ1 and RQ2, we conducted the first developer study, which sheds light on how the developers of the Solana ecosystem handle security and the corresponding challenges they face. Our study comprised a 90-minute code review study with 35~participants and follow-up interviews with a subset of seven~participants. We asked participants to write code reviews for a smart contract split into three parts, 
each involving one of the most common~\cite{vrust} types of security vulnerability (Missing Signer Check, Integer Bugs, and Arbitrary Cross-Program Invocation). Participants assessed their knowledge about these vulnerabilities and explained what challenges they pose in the context of securing smart contracts. 
Our analysis showed that none of the participants spotted all vulnerabilities in the code review tasks despite their claimed confidence in addressing them. Additionally, participants referred to a shortage of qualified Solana developers, leading to the hiring of inexperienced individuals. The lack of documentation, code reviews, audits, testing, and the complexity of Rust was reported as the main challenges for developers, leading them to adopt alternative frameworks such as Anchor~\cite{anchor}.

\paragraph{Impact Analysis}
Given these results, one would expect that the investigation of RQ3 would lead to a highly vulnerable ecosystem. 
However, recent studies~\cite{smolka2023fuzz} showed that only \pnum{52}{6324} projects were found to be vulnerable in Solana. 
To further confirm those results, we built a framework using symbolic execution to detect the Arbitrary Cross-Program Invocation~(\acpi) vulnerabilities in currently deployed Solana smart contracts. These vulnerabilities were among the most challenging to detect by developers, according to our study. Using our tool, we then automatically analyzed all 6,324 smart contracts deployed on Solana. Fortunately, our findings corroborate the results~\cite{smolka2023fuzz} and show that only \pnum{14}{6324} deployed smart contracts are vulnerable to ACPI.

Our analysis suggests two conflicting results. On the one hand, Solana developers do not seem to have domain expertise in addressing security challenges when developing smart contracts. On the other hand, the prevalence of security vulnerabilities is fortunately not severe in current Solana smart contracts. Our developer study suggests, however, that the popular Anchor framework provides a healthy tooling environment for developers and is probably one of the most important reasons for the low prevalence of vulnerabilities in the Solana smart contract ecosystem. In fact, our analysis shows that more than 88\% of existing Solana smart contracts are currently developed with the help of the Anchor framework (and not with direct native Rust support). We, therefore, hope that our findings motivate further research highlighting the importance of frameworks, such as Anchor, in aiding the development of secure smart contracts.


\section{Background}
\label{sec:SolanaProgramVulnerabilities}

Solana is a high-performance smart-contract enabled blockchain platform designed to address the typical scalability and throughput challenges faced by traditional blockchain networks~\cite{solwhitepaper}.

Solana relies on \emph{accounts} as the primitive data structure to store data on the blockchain (see \Cref{fig:sol_arch}). They are used for almost all kinds of storage, including user wallets, the contract code, and state.
\begin{enumerate*}[(1)]
\item User wallets are accounts held by off-chain entities, authorized via a key pair,
\item data accounts are \emph{owned} by smart contracts and can store arbitrary, contract-related information, and
\item smart contract accounts  are marked as executable and store the bytecode of the contract.
\end{enumerate*}
All of these accounts have a key, an owner, and store a balance of Solana's default currency (Sol/Lamports), but only data accounts can store additional arbitrary state (for contracts). Due to lack of space, we provide additional details about Solana in \Cref{sec:solana-background}.

We now detail the vulnerabilities that were selected for the user study. 
In our study, we focused on investigating the 3~most prominent vulnerabilities outlined by prior work, namely integer bugs, arbitrary cross-program invocation, and missing signer check, which account for 48\%, 16\%, and 8\%, respectively, of all vulnerabilities in available smart contracts according to~\cite{vrust}. In a more recent study~\cite{smolka2023fuzz}, it was reported that these vulnerabilities account for up to 93\% (36\%, 36\%, 21\%, respectively) of all vulnerabilities in Solana smart contracts. Note that integer bugs are of particular interest due to their high prevalence, even though Rust provides an automated overflow/underflow check that can be enabled at compile time  (i.e., setting $\emph{overflow-checks} = \emph{true}$ in the Cargo.toml file)\footnote{Surprisingly, this option seems to go largely unnoticed by Solana developers and is not mentioned in the Solana documentation.} to effectively mitigate this vulnerability.

\paragraph{Missing Signer Check (\msc)}
Recall that, since \ssc are stateless, all information on the current state of the contract has to be provided as input of the transaction.
When the user wants to access a data account (e.g., their vault), the smart contract has to verify that the account matches the user wallet to avoid unauthorized access.
However, developers additionally need to ensure that the user wallet provided as input has signed the transaction.
Smart contracts that are missing proper signer validation would allow a malicious user to claim a different user's data as their own and effectively be able to use their tokens.

\paragraph{Integer Bugs (\iobug)}
\iobug occur when the result of an arithmetic operation grows either greater (integer overflow) or lower (integer underflow) than the size of the target type allows~\cite{io_bugs_sc}.
Unlike other programming languages, Rust adds runtime checks for integer overflows and underflows in its \emph{dev} mode, typically used during code development and debugging.
However, when building in release mode, the default behavior of Rust does not include such checks on arithmetic operations for performance optimizations---opening the door for \iobug~\cite{rustprofile}.
Well-known \iobug on smart contract platforms include the BEC-token hack that was actively exploited through an unsafe multiplication, resulting in a loss of $10^{58}$~BEC tokens~\cite{bec-report}, each worth around $0.30$~USD at the time of the hack.

\paragraph{Arbitrary Cross-Program Invocation (\acpi)}
Cross-program invocation~(CPI) allows \ssc to interact with other \ssc through atomic operations.
This feature enables complex and decentralized applications to collaborate and compose seamlessly within the Solana ecosystem.
\acpi occurs when a program calls into an untrusted/user-provided program, leading to unexpected behavior and unauthorized access to data, resulting in the possibility for the adversary to modify the victim contract's data.
The typical way to prevent \acpi is to ensure that a smart contract only performs cross-program invocation to known or trusted addresses.
In order to improve the security of \ssc, the Solana community developed the Anchor framework~\cite{anchor} that automates owner and signer checks.

\section{Methodology}
To investigate if Solana smart contract developers recognize  security vulnerabilities and the security challenges they face, 
we conducted a code review study with 35 participants and semi-structured interviews with a subset of seven~participants.
Participants were invited to a 90-minutes online study to review three vulnerable files of a simple smart contract written in Rust.
To avoid priming participants for security, we asked them to write code reviews as usual without mentioning security.
After completing the review task, participants were asked to fill out a questionnaire on their task experience and demographics, and were invited to a follow-up interview.
A detailed overview of our study procedure can be found in \Cref{fig:study_strucsture} and the replication package is available in \Cref{sec:replication}.

\subsection{Code Review Study}
\label{subsec:methodology_survey}
\begin{figure}[t]
    \centering
    \includegraphics[bb=0 0 180 265, width=.6\linewidth]{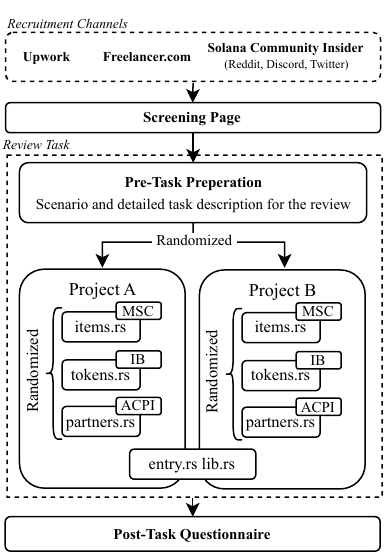}
    \caption{Structure of the Code Review Study}
    \label{fig:study_strucsture}
\end{figure}

For the code review task, we focused on the security vulnerabilities described in \Cref{sec:SolanaProgramVulnerabilities}. We embedded the \iobug, \msc, and \acpi vulnerability each in one file of a smart contract (see \Cref{sec:StudySmartContract}). 
Further, we added multiple non-security-related issues in the smart contract (distraction tasks) to make the review task more realistic and security issues less obvious. 
To mitigate potential ordering effects, the three files were shown randomly to participants.
Additionally, we created a second smart contract (see Project~A and~B in \Cref{fig:study_strucsture}), where we changed the order of the security vulnerabilities and distraction tasks.
We created a GitLab snippet for each relevant file of the two contracts and embedded each on a separate page. Participants were asked for a code review by providing comments and fixes.
Participants were presented with a table containing three columns: "Line" to input the line of code, "Comments" to record comments for the specific line, and "Code Suggestion" to propose solutions.
Further, they were asked whether the current state of the code is \emph{releasable without any risks}, \emph{releasable with minor risks}, or whether the code is \emph{not releasable}, and to explain their decision. 

\subsubsection{The Study's Smart Contract}
\label{sec:StudySmartContract}%
\newcommand\price{\texttt{PRICE}\xspace}%
\newcommand\varcount{\texttt{count}\xspace}%
\newcommand\amt{\texttt{amount}\xspace}%
We built a small marketplace smart contract in Rust that allows users to buy a newly defined token at a fixed price and exchange them among each other. It is also possible to own and exchange items, where buyers could send offers to buy a specific item, and the seller could select the most appropriate offer to sell their item for.
The marketplace smart contract was split across three~main files that tackle three~different parts of the marketplace. 
The files were used for the review task, each containing one type of vulnerability and one or more distractions depending on the file size. The complete source code of the marketplace is part of our replication package Appendix~\ref{sec:replication} and a summary of the files is included in \Cref{tab:task-vulns}.

\begin{table}[H]
\caption{Vulnerabilities and non-security-related issues (distractions) of the review task.}
\label{tab:task-vulns}
\centering
\begin{adjustbox}{max width=\linewidth}
\begin{tabular}{@{}lccccc@{}}
    \toprule
    \multirow{2}{*}{\textbf{File}} & \clap{\textbf{Vulnerability}} &\multicolumn{2}{c}{\textbf{Line number}} & \textbf{Number of} & \textbf{Lines} \\
    & \textbf{Type} & \textbf{A} & \textbf{B} & \textbf{Distractions} & \textbf{of Code} \\
    \midrule
    \texttt{item.rs} & \msc & 53--77 & 84--108 & 2 & 120\\
    \texttt{tokens.rs} & \iobug & 95 \& 126 & 95 \& 156 & 1 & 140\\
    \texttt{partner.rs} & \acpi & 52 & 73 & 3 & 60\\
    \bottomrule
\end{tabular}
\end{adjustbox}
\end{table}
\paragraph{tokens.rs}
The file \texttt{tokens.rs} defines the tokens used in our marketplace and its main functionalities, such as selling and buying tokens against Lamports, distributing tokens to multiple accounts, and peer-to-peer transfer of tokens. In this file, we inserted \iobug in the distribute function similarly to the vulnerability of the BEC token on Ethereum~\cite{bec-report}. The function takes as input an \amt and a list of accounts and sends \amt of token to each recipient account in the list. The file contains 140 lines of code.
To avoid overdrafting an account, the smart contract does verify that the amount owned is bigger than the total to send, as shown in this  code snippet:
\begin{minted}{rust}
let count = account.len() - 3;
...
let total = amount * (count as u64);|\label{lst:iobec}|
assert!(token_vault.amount >= total);|\label{lst:iobecassert}|
...
for receiver_info in receiver_infos {
    ...
    receiver.amount += amount;|\label{lst:fakeiob}|
}

\end{minted}

Since the first three accounts are the user wallet and two data accounts of the smart contract, the number of accounts to distribute to is the length of the account list~$-~3$.
As seen in line~\ref{lst:iobec}, if $\amt \times \varcount$ overflows, one can send a large amount of tokens while bypassing the overdraft check of line~\ref{lst:iobecassert}.
Another potential overflow can be flagged in line~\ref{lst:fakeiob}, where sending a large amount of tokens to a user could cause an overflow and token losses.
However, this overflow can only be exploited by first exploiting the overflow in line~\ref{lst:iobec} as otherwise it is not possible to get enough tokens to overflow to recipient: the smart contract guarantees that the (legitimate) total supply amount never exceeds the capacity of a \texttt{u64} variable, preventing any overflow caused by a legitimate total in supply.
We also introduced a second, more covert, \iobug vulnerability in the buy function that takes as input the amount of tokens to buy:
\begin{minted}{rust}
const PRICE: u64 = 1000000;  // 0.001 SOL
...
invoke(
    &system_instruction::transfer(authority_info.key,
            reserve_info.key, amount * PRICE),|\label{lst:iobug}|
    &[authority_info.clone(), vault_info.clone()],
)?;
\end{minted}
In this case, the \price is hardcoded and fixed to 0.001 SOL. However, since the amount is user provided, a malicious user could carefully select a value of \amt such that the operation $\amt \times \price$ in line~\ref{lst:iobug} overflows to 1 (or a small value) SOL, and pay only 1 SOL for a huge amount of tokens. This \iobug is less obvious due to the multiplication with a constant number.
Due to the already large file size, we only included one distraction: one of the function of the smart contract uses bad variable names (single letter variables). 

\paragraph{item.rs}
We define the logic to exchange items in the file \texttt{item.rs}. Here, users can place, accept, and delete offers for existing items for trading them against our defined token. The file contains 120 lines of code and we included an \msc vulnerability. The following is a simplified snippet of the vulnerable code:


\begin{minted}[samepage]{rust}
let offer_info     = next_account_info(acc_info_iter)?;
let item_info      = next_account_info(acc_info_iter)?;
let vault_info     = next_account_info(acc_info_iter)?;
let user_info      = next_account_info(acc_info_iter)?;
let authority_info = next_account_info(acc_info_iter)?;

let mut offer     = Offer::try_from_slice(&offer_info.data)?;
let (offer_key, _) = Pubkey::find_program_address(...);
let mut item      = Item::try_from_slice(&item_info.data)?;
let user          = User::try_from_slice(&user_info.data)?;
let mut vault
    = TokenVault::try_from_slice(&vault_info.data)?;

assert_eq!(offer_key, *offer_info.key);|\label{lst:offerkey}|
assert_eq!(offer.item, *item_info.key);|\label{lst:itemkey}|
assert_eq!(item.user, *user_info.key);|\label{lst:userkey}|
assert_eq!(vault.user, *user_info.key);|\label{lst:vaultkey}|
assert_eq!(user.authority, *authority_info.key);|\label{lst:authoritykey}|

item.user = offer.from;
vault.amount += offer.amount;
\end{minted}

Here, the code ensures that
\begin{enumerate*}[(i)]
\item the offer key is valid (line~\ref{lst:offerkey}) and references the given item (line~\ref{lst:itemkey}) to avoid selling an unwanted item to an offer,
\item the item is owned by the user (line~\ref{lst:userkey}) to prevent the selling of an item you do not own,
\item the vault is owned by the user (line~\ref{lst:vaultkey}) to ensure the tokens will be credited to the correct account, and
\item the user account is owned by the user wallet (line~\ref{lst:authoritykey}).
\end{enumerate*}
However, there is no verification that the user wallet is \emph{signer}, and hence an attacker could force anyone to accept their offer, regardless of the actual offer, effectively allowing them to steal the items of other users.
We included two distractions in this file, namely bad variable naming and excessively long line lengths (more than 280 characters). 

\bgroup
\newcommand\is{\texttt{invoke\_signed}\xspace}
\newcommand\ci{\texttt{coupon\_info}\xspace}
\newcommand\pp{\texttt{partner\_prog}\xspace}
\newcommand\rc{\texttt{refresh\_credit}\xspace}

\paragraph{partner.rs} In the last file \texttt{partner.rs}, we implemented a functionality that allows users to redeem "coupons" sponsored by "partners".
Partners hand out coupons to users and users redeem them in this smart contract to gain tokens. The code snippet of the vulnerable function is shown in \Cref{fig:marketplace-cpi}.

\begin{figure}[b]
\begin{minted}{rust}
pub fn refresh_credit(program_id: &Pubkey, 
  accounts: &[AccountInfo], seed: String) -> ProgramResult {

  // Read two accounts
  let account_info_iter = &mut accounts.iter();
  let coupon_info = next_account_info(account_info_iter)?;
  let partner_prog = next_account_info(account_info_iter)?;

  let (coupon_key, bump) = Pubkey::find_program_address(
    &[program_id.as_ref(), seed.as_bytes()], program_id);

  // Validate coupon owner
  assert_eq!(*coupon_info.key, coupon_key);
  assert_eq!(coupon_info.owner, program_id);
  assert!(partner_prog.executable);

  // Invoke partner_prog to refresh credits
  let instruction = Instruction::new_with_borsh(
    *partner_prog.key, // Instruction target not validated
    &ExtPartnerInstructions::RefreshCredit,
    vec![AccountMeta::new(coupon_key, true)]
  );
  invoke_signed(&instruction, &[coupon_info.clone()],
    &[&[program_id.as_ref(), seed.as_bytes(), &[bump]]]
  )
}
\end{minted}
\caption{ACPI used in the study.}
\label{fig:marketplace-cpi}
\end{figure}

The \rc function is responsible for requesting the current coupon value in tokens from the respective partner.
This must be done before the user can redeem the coupon to obtain the tokens.
The actual update of the coupon value is implemented in the smart contract of the partner, which is called using \is.
Note that \is gives the called contract the privileges to act on the callers' behalf.
Thus, the usage of \is must be carefully reviewed to avoid vulnerabilities.
In this case, \is is required to enable the \pp to update the value in \ci.
However, the \rc function does not validate that the \pp corresponds to the \ci, thus, allows an attacker to call \emph{any} smart contract.
An attacker can exploit this lack of validation by using a crafted \pp that populates the \ci with arbitrary data, e.g., a value of millions of tokens.

Since this file is much smaller compared to other files (only 60 lines of code), we introduced three distractions, namely bad variable naming and exceeding line length similar to \texttt{item.rs} as well as variable shadowing, where a variable name is reused. This is typically considered bad practice as it can be difficult to notice the variable was redeclared and changed type.
\egroup

\subsubsection{Post-Task Questionnaire}
To ensure data quality, participants were shown an attention check question after task completion and were asked to fill out a \emph{Post-Tasks Questionnaire} (see Appendix~\ref{sec:appendix_posttask}). We asked how they reviewed the code (e.g., if a guideline or tool was used, if they usually consult a colleague for such tasks, etc.) and how much time they dedicate to security. We were also interested in whether they spend more, less, or equal time on security when using Rust compared to other programming languages. We asked how confident they are dealing with security issues while developing Solana smart contracts and what they think about why vulnerabilities exist. Finally, we collected participants' demographics and payment details and invited them to a follow-up interview.

\subsubsection{Pilot Study}
We tested our study with four participants. The first two participants were researchers familiar with Rust and Solana smart contracts in an academic context. Further, we recruited two Solana Smart contract developers through our professional contacts in the Solana community. We received feedback that understanding the code might be challenging for smart contract developers. Therefore, we added explanation comments to our code.

\subsection{Interview Study}

After completing the code review study, participants were invited to a follow-up semi-structured interview lasting 45~minutes. All the interviews were conducted online in English between March and April 2023 using Zoom~\cite{zoom}. 
They were audio recorded and transcribed.
Two authors were present during all interviews, with one interviewer and one assistant taking notes and providing supplementary questions. 
The interview guideline covered five main themes concerning Solana smart contract development:
\begin{enumerate*}[(1)]
\item The participants' role and experience,
\item the process,
\item vulnerabilities existing,
\item Rust safety, and
\item suggestions for security improvement.
\end{enumerate*} 
To test the guideline, we conducted a pilot study with two researchers experienced in Solana smart contract development. Their feedback led to minor 
changes.
The final interview guideline can be found in Appendix~\ref{sec:appendix_interview_guide}.

\subsection{Recruitment}
\label{sec:participants}
We used different channels to recruit participants experienced in developing Solana smart contracts using Rust. The participants qualified if they were
\begin{enumerate*}[(1)]
\item experienced in developing Solana smart contracts using Rust,
\item willing to perform a code review task,
\item older than 18~years, and
\item feeling comfortable filling out a survey in English.
\end{enumerate*}
We started recruitment by using our professional contacts in the core Solana community.
Our study was advertised by the core Solana community on platforms like Twitter and Reddit and spread within the Solana community (e.g., internal Discord).
Additionally, we recruited participants on the freelancing platforms \emph{Upwork.com} and \emph{Freelancer.com}.
For Upwork, we manually evaluated each interested participant’s account to ensure they were experienced in Rust and Solana.
Further, we used the platform’s search tool to identify potential candidates based on the tags \emph{Solana} and \emph{Rust}.
We manually checked their profiles, as the tag search often returned false positives.
After contacting 110~participants on Upwork, we stopped recruiting as the Upwork search did not provide new candidates matching our criteria. Of the initial 110 participants, 25 actively participated in the study, with 22 completing the survey.

On \emph{Freelancer.com}, we reached out to approximately 15~freelancers using the platform's search engine and stopped recruiting when the search engine did not show new results searching for Solana and Rust expertise. We received five replies to our first study advertisement post.
Three of them sent generic messages and 
one expressed interest but was not invited to the study because they had no prior experience in developing \ssc with Rust.
Only one participant accepted our invitation and completed the survey.
We also reached out to three \emph{LinkedIn.com} blockchain groups but did not receive feedback from the moderators.
Thus, we could not recruit participants through this channel.
We stopped recruiting for our study when no more new participants joined, and we ran out of further recruitment channel options. Despite extensive outreach, participant recruitment proved challenging.
A summary of participants recruited per channel is provided in \Cref{tab:partChannel}.

\begin{table*}[ht]
\caption{Summary of participant recruitment per channel.}
\label{tab:partChannel}
\centering
\scriptsize
\renewcommand{\arraystretch}{1.1}
\setlength{\tabcolsep}{0.6\tabcolsep}
\setlength{\defaultaddspace}{0.33\defaultaddspace}
\begin{tabular}{lccccc}
    \toprule
    \textbf{Survey Channel} & \textbf{\# of Started Survey} & \textbf{\# of Finished Survey} & \textbf{\# of Valid Responses} & \textbf{\# of Spammers} & \textbf{\# of Bots} \\
    \midrule
    \parbox{5cm}{\textbf{Solana community insider\\\null\quad(Reddit, Twitter, Private Discord Channels)}} & 230 & 32 & 15 & 11 & 6 \\
    \textbf{Upwork} & 25 & 22 & 19 & 3 & 0 \\
    \textbf{Freelancer} & 1 & 1 & 1 & 0 & 0 \\
    \midrule
    \textbf{Total} & \textbf{256} & \textbf{55} & \textbf{35} & \textbf{14} & \textbf{6} \\
    \bottomrule
\end{tabular}
\end{table*}

\subsection{Participants}
While we reached 256~participants across all channels, 55~completed our survey. 
To ensure data quality, two researchers conducted a thorough manual analysis of each survey response, focusing on
\begin{enumerate*}[(1)]
\item assessing the quality and relevance of the answers to the survey questions,
\item evaluating the participants' years of experience with developing and reviewing Solana smart contracts,
\item tracking the time taken to complete the three code review tasks,
\item identifying responses flagged as potential bots by the used survey platform, Qualtrics~\cite{qualtrics}
\end{enumerate*}.
Our manual analysis identified 20~participants as spammers out of the 55~participants. Six of these 20~participants were also identified as bots by our survey platform Qualtrics. These spammers/bots had low-quality responses, such as providing the same response for every open-ended question in the survey or leaving irrelevant comments in the code review section.
For example, a participant mentioned: \simplequote{There's one row in the table for each SQL statement that you run and each row is uniquely identified by the SQL\_ID column. You can use this SQL\_ID to track a given SQL statement throughout the Oracle database,} although the code snippets had no relation to SQL.

It is important to note that Solana has existed since March 2020~\cite{Adams}, making it barely more than three~years old at the time of the study. Thus, participants were excluded if claiming to have unrealistically long experience, such as 12~years of experience reviewing Solana smart contracts or six~years developing Solana smart contracts. We excluded one participant, taking one and a half minutes to review all three files.  

In total, we excluded 20~participants from data analysis due to quality issues leaving us with 35~valid participants' answers. 
Note there are approximately 2000~Solana developers in 2022, representing about \p{1.75} of the total population~\cite{stepfinance, Melinek_2023}.
Thirteen participants were interested in the follow-up interview, with seven~accepting our invitation. Our participants were from 15 different predominantly non-western countries. Four participants did not want to disclose their age. More than two-thirds (\p{77.4}) of our participants reported that their age were below~30. \pnum{31}{34} had at least a Bachelor’s degree. The average experience as a software developer was 4.16~years but varied (min. one~year, max. 10~years). 31~participants had at least one year of experience with Rust, and 16 had at least two years of experience with Solana (Solana started around March 2020~\cite{Adams}).
\Cref{tab:demographics} provides an overview of our participants’ demographics. A detailed table on the demographics of the interviewed participants can be found in \Cref{tab:participants}.
All the participants were compensated with 50\,\$ for the code reviewing task and an additional 50\,\$ for the follow-up interview. Participants from non-freelancer platforms could choose between compensation via PayPal or an Amazon voucher. Participants were asked about the fairness of the compensation amount, and the majority of participants \pnum{29}{35} stated that the payment was "just right".


\begin{table}[tb]
\caption{Demographics of 35 participants.}
\label{tab:demographics}
\centering
\begin{adjustbox}{max width=\linewidth}
\scriptsize
\setlength{\tabcolsep}{0.66\tabcolsep}
\setlength{\defaultaddspace}{0.25\defaultaddspace} 
\begin{threeparttable}
\begin{tabular}{@{}llrrlrr@{}}
    \toprule
    \multicolumn{7}{@{}l}{\textbf{Gender}} \\
    & Male & 30 & \p{85.7} & Prefer not to disclose & 3& \p{8.6} \\
    & Female & 2 & \p{5.7} \\

    \addlinespace
    \multicolumn{7}{@{}l}{\textbf{Countries}} \\
    & India & 7 & \p{20.6} & Ukraine & 5 & \p{14.7} \\
    & Pakistan & 3 & \p{8.8} & Georgia & 2 & \p{5.9} \\
    & UK & 2 & \p{5.9} & Nigeria & 2 & \p{5.9} \\
    & US & 2 & \p{5.9} & Prefer not to disclose & 3 & \p{8.8} \\
    & Other\tnote{\textdagger} & 8 & \p{23.2} \\
    \addlinespace
    \multicolumn{7}{@{}l}{\textbf{Age [years]}} \\
    & Min. & 19 & & Max. & 42 & \\
    & Mean (Std.) & 25.19 & $\pm$6.12 & Median & 23.0 \\
    \addlinespace
    \multicolumn{7}{@{}l}{\textbf{Industry Experience [years]}}\\
    & Min. & 0.0 & & Max. & 11 & \\
    & Mean (Std.) & 4.36 & $\pm$3.13 & Median & 3.0 \\
    \addlinespace
    \multicolumn{7}{@{}l}{\textbf{Software Developer Experience [years]}}\\
    & Min. & 1.0 & & Max. & 10 & \\
    & Mean (Std.) & 4.16 & $\pm$2.97 & Median & 3.0 \\
    \addlinespace
    \multicolumn{7}{@{}l}{\textbf{Solana Experience [years]}}\\
    & Min. & 0.3 & & Max. & 4 & \\
    & Mean (Std.) & 1.59 & $\pm$0.86 & Median & 1.5 \\
    \addlinespace
    \multicolumn{7}{@{}l}{\textbf{Solana Review Experience [years]}}\\
    & Min. & 0.0 & & Max. & 4 & \\
    & Mean (Std.) & 1.28 & $\pm$0.97 & Median & 1.0 \\
    \addlinespace
    \multicolumn{7}{@{}l}{\textbf{Rust Experience [years]}}\\
    & Min. & 0.3 & & Max. & 7 & \\
    & Mean (Std.) & 1.99 & $\pm$1.47 & Median & 2.0 \\
    \addlinespace
    \multicolumn{7}{@{}l}{\textbf{Education}} \\
    & Bachelor's degree & 24 & \p{68.6} & Graduate school & 2 & \p{5.7} \\
    & High school or equivalent & 2 & \p{5.7} & Master's degree & 4 & \p{11.4}\\
    & Vocational degree & 1 & \p{2.9} & Doctorate/PhD & 1 & \p{2.9} \\
    & Prefer not to disclose & 1 & \p{2.9} \\
    \bottomrule
\end{tabular}
\begin{tablenotes}
    \footnotesize
    \item[\textdagger] Each country occurring once.
\end{tablenotes}
\end{threeparttable}
\end{adjustbox}
\end{table}

\subsection{Analysis}
\paragraph{Code Review Study} 
We were interested in the results of the reviews if
\begin{enumerate*}[(1)]
\item security vulnerabilities were found,
\item the given fixes were correct,
\item the distractions were found,
\item erroneous vulnerabilities were found, or
\item improvements were suggested.
\end{enumerate*}
We have, therefore, opted for a deductive approach by coding the comments and fixes.
Two researchers coded all statements in collaborative coding sessions.
Open survey questions were coded collaboratively by following the same coding process as the interviews. 

\paragraph{Interviews}
The interview transcripts were analyzed using thematic analysis, referring to~\citewauthor{kuckartz2013qualitative} as a guide for creating our codebook.
The coding process was conducted using the qualitative analysis tool MAXQDA~\cite{maxqda}. Two researchers independently coded the seven interviews and developed a code system based on the interview data. Following the consensual coding process outlined by~\citewauthor{hopf1993verhaltnis}, the two researchers met after completing the initial coding round to compare and discuss individual coding results. To ensure a shared understanding of the coding categories, definitions, and interpretation, discrepancies or disagreements in coding were resolved through discussion, reaching full consensus. 
Based on the discussions, the codebook was refined and all data has been recoded afterwards again. 
We assigned 392 codings with a median of 50 codings per interview. The final codebook can be found in the appendix (\Cref{tab:codebook}).

\subsection{Limitations}

We note that our study's findings may not be generalizable, and more research is needed to explore the security challenges faced by \ssc developers. Despite using various recruitment channels, we fell short of our target for Solana developers and observed limited representation from Western countries. 
During our study, we encountered \pnum{5}{35} participants who reviewed the Solana smart contracts differently than expected despite claiming prior reviewing experience. However, they analyzed the code, tried to understand it, shared their thoughts on security challenges, and decided whether it should be released. We opted to include these participants since (1) their survey responses provided valuable insights, and (2) this might reflect real-world scenarios.
We stress that our results remain consistent even if we exclude these five participants; namely, the share of participants that did not find any vulnerability would change from \pnum{28}{35} to \pnum{23}{30}; not causing a significant impact on the outcome of the study (see \Cref{tab:release_answers}).

\subsection{Ethics and Data Protection}
The study received approval from our Institutional Review Board~(IRB). 
Participants were presented with a consent form informing them about the study's purpose, data usage, duration, and associated risks. We complied with the 
General Data Protection Regulation~(GDPR). 
Participants were informed that they could opt out at any time without any negative consequences. 
and were assured that only pseudonymized data and quotes would be published.
After task completion, participants were debriefed that we sought to explore various security challenges witnessed in the development of Solana smart contracts in our study.
During our study, we flagged 20 participants as spammers. For 16, we were able to determine this prior to compensation. To ensure that our analysis was correct, we informed these participants that their results were flagged as spam and offered them a second chance to complete the survey if they wished to receive compensation. 
However, our system again flagged all their re-submission clearly as spam. Due to limited research funding and high compensation for the study, we contacted the IRB board to inquire about compensating these participants. An agreement was made with the IRB to not compensate these participants with the hope of discouraging future research fraud. While past research studies did not report such issues with ``general'' software developers, it seems more preventive measures are specifically required for recruitment and data collection in online studies with Solana smart contract developers.


\section{Study Results}
\label{sec:study-results}
In this section, we present the results of the code review study and the follow-up interviews.

\subsection{Code Review Results}
\label{sec:code_review_results}
We received 253 comments and 146 fixes from 35 participants for the code reviews. On average, each participant reported 7.23 comments with a standard derivation of 6.22. Five participants did not report any comment at all. However, they still provided reasons for their release decisions (e.g., \surveyquote[P9]{9}{Reviewed the code and it seems good}). On average, participants took 68.84 (median 52.19) minutes to review all three files of one smart contract with a standard deviation of 70.80 minutes and a median of 52.79 (ranging from 7.15 to 378.94 minutes). 
We also asked participants to report the average time they needed to review one file. \Cref{tab:avg_timings} compares the time we measured with the self-reported statements. \pnum{10}{35} participants stated to need (at least \p{250}) more and \pnum{16}{35} participants less time than we measured. However, all people passed our attention check question.
\pnum{24}{35} participants provided at least one comment that (1)~was not understandable, (2)~did not provide any insights, or (3)~was incorrect. In addition to correct and incorrect reported weaknesses, incorrect suggestions, and distractions, \pnum{9}{35} participants also reported improvements (e.g., suggestions to reduce the gas fee). 
We note that \pnum{5}{35} participants reviewed the Solana smart contracts differently than expected despite claiming prior experience reviewing such contracts. Their code review approach consisted of analyzing the code to comprehend its functionality. Further, these participants claimed to focus on code readability, best practices, and security vulnerabilities during their review. They also assessed the readiness for the release of the given smart contracts.


\begin{table}[t]
    \caption{ Self-reported and actual measured average time of participants per task.}
    \label{tab:avg_timings}
    \centering

\begin{adjustbox}{max width=\linewidth}
\begin{tabular}{@{}lccccc@{}}
    \toprule
    \textbf{}                        & \textbf{Mean} & \textbf{Median} & \textbf{Std} & \textbf{Max} & \textbf{Min} \\ \midrule
    Self reported avg. time per task & 28.94         & 30.0            & 22.46        & 120.00       & 3.00         \\
    Measured avg. time per task      & 24.48         & 17.6            & 23.36        & 126.31       & 3.17         \\ \bottomrule
\end{tabular}
\end{adjustbox}
\end{table}

\subsubsection{Vulnerabilities Found} 
None of our participants found all the security vulnerabilities in the smart contracts, and not one participant spotted both relevant \iobug vulnerabilities within \emph{tokens.rs}. 
\Cref{tab:review_results} provides an overview of the security vulnerabilities or distractions found by participants within the related files. Four participants (\p{11.4}) spotted at least one of the \iobug vulnerabilities, four~(\p{11.4}) identified the \msc vulnerability and two~(\p{5.7}) found the \acpi vulnerability.
Despite Rust providing automatic prevention for \iobug, none of the participants mentioned this option when proposing solutions for the \iobug they discovered. 
The bottom of \Cref{tab:review_results} presents the number of participants who were able to find multiple valid vulnerabilities or distractions. Only a few participants found vulnerabilities in more than one file. Furthermore, no participant was able to spot all distractions in all files. Only five participants (\p{14.3}) found at least one distraction. Although the level of detail varied among the provided fixes, all fixes by those who identified a vulnerability were correct, except one participant did not provide a fix for the \acpi vulnerability. 


\begin{table}[tb]
    \caption{Number of participants found distractions and security vulnerabilities.}
\label{tab:review_results}
    \centering
    \begin{adjustbox}{max width=\linewidth}
   \begin{tabular}{@{}ll@{}cccc@{}}
\toprule
\multirow{2}{*}{\textbf{Vulnarability type}}                & \multirow{2}{*}{\textbf{File}}     & \multicolumn{2}{c}{\textbf{Distractions}} & \multicolumn{2}{c}{\textbf{Vulnerabilities}} \\
                                                              &                                    & \textbf{\# Found > 0}  & \textbf{\# Found all}  & \textbf{\# Found > 0}    & \textbf{\# Found all}   \\ \midrule
\msc                                           & items.rs                           & 3                         & 1             & 4                           & 4              \\
\iobug                                         & tokens.rs                          & 3                         & 3             & 4                           & 0              \\
\acpi                                          & partners.rs                        & 2                         & 0             & 2                           & 2              \\ \midrule
\msc \& \iobug                  & items.rs \& tokens.rs              & 2                         & 1             & 2                           & 0              \\
\msc \& \acpi                   & items.rs \& partners.rs            & 1                         & 0             & 1                           & 1              \\
\iobug \& \acpi                 & tokens.rs \& partners.rs           & 1                         & 0             & 1                           & 0              \\
\msc \& \iobug \& \msc & all three & 1                         & 0             & 1                           & 0              \\ \bottomrule
\end{tabular}
    \end{adjustbox}
\end{table}

Except for the participant that did not provide a fix for for the identified \acpi vulnerability, all others who found a vulnerability in one of the files, opted for the option to not release the code. 
\Cref{tab:release_answers} further shows the number of release decisions made by the participants.
We also coded if participants reported erroneous security vulnerabilities. Five participants~(\p{14.3}) reported false security vulnerabilities in \emph{items.rs} (\msc), four~(\p{11.4}) in \emph{tokens.rs} (\iobug) and no false positives have been reported for \emph{partners.rs} (\acpi). Seven participants (\p{20.0}) found at least one valid vulnerability, three~(\p{8.6}) found at least two vulnerabilities and only two~(\p{5.7}) found more than two vulnerabilities in all files. Moreover, our results indicate that the participants found all three investigated vulnerabilities equally challenging to detect.

We also cross-checked whether there is a link between the number of vulnerabilities found and their experience in performing code reviews. To do so, we computed a Pearson correlation coefficient to assess the linear relationship between \emph{the total vulnerabilities found by each participant} and \emph{(a) the participants' years of experience in reviewing Solana smart contracts}, as well as \emph{(b) the total number of Solana smart contracts they have reviewed}. Both variables showed very weak negative correlations (a: r=-0.049, p=0.78; b: r=-0.0218, p=0.90) and were not significant.


\begin{table}[tb]
    \caption{Code review outcomes. \\ {\footnotesize Numbers in round brackets show the answers of those who found a valid vulnerability for that file,~while~numbers in curly brackets exclude the five participants that misunderstood the task.} }
    \label{tab:release_answers}
    \centering
   \begin{adjustbox}{max width=\linewidth}
   \begin{tabular}{@{}lllll@{}}
\toprule
\textbf{Vulnerability Type} & \textbf{File} & \textbf{Releaseable} &\parbox{2.3cm}{\textbf{Releasable with \\minor risks}} & \textbf{Not releaseable} \\ \midrule
\msc         & items.rs      & 11 {\{8\}} & 12  {\{10\}} & 12 (4)                   \\
\iobug       & tokens.rs     & 14 {\{11\}} & 12  {\{10\}} & \phantom{1}9 (4)                    \\
\acpi        & partners.rs   & 14 {\{9\}} & 10 (1)                               & 11 (1)                   \\ \bottomrule
\end{tabular}
    \end{adjustbox}
\end{table}

\subsubsection{Knowledge of Vulnerabilities} 
In the survey, participants were asked to rate their confidence in explaining Solana security vulnerabilities and ways to mitigate them. The majority of participants rated themselves as completely or fairly confident in explaining and addressing certain vulnerabilities. including missing ownership \pnum{26}{35}, \msc \pnum{25}{35} and integer bugs \pnum{23}{35}. However, confidence in the knowledge of \acpi was lower, with only \pnum{18}{35} feeling completely or fairly confident.

To further assess participants' knowledge, during the interviews, we focused on their understanding of the three specific security vulnerabilities that were present in the code review tasks.
Six participants demonstrated confidence in explaining \msc, while they showed less confidence in explaining integer bugs. However, participants struggled the most when it came to explaining \acpi, with only three participants familiar with this vulnerability. In comparison to the self-ratings in the survey, five interview participants rated themselves relatively fairly, with some even underestimating their own ability. Notably, only two participants showed a slight over-confidence with their self-rating.

\begin{summary}{RQ 1 -- Summary}
None of our participants spotted all vulnerabilities and distractions in the code review task, although most claimed to be confident in explaining and mitigating these vulnerabilities. Only seven participants (\p[35]{7}) have found at least one vulnerability and 6 provided a proper fix. It is alarming that \pnum{29}{35} would have released at least one vulnerable smart contract source code file.
\end{summary}

\subsection{Developers' Security Challenges}
\label{sec:Security_Challenges}
We now present our results from the code review study and the interviews.
We identified six security challenges encountered by developers during the development of Solana smart contracts.
In \Cref{tab:vuln_Challenges}, we provide participant numbers from the survey to demonstrate the frequency and distribution of challenges.

In the following, we provide deeper insights into the identified challenges.

\cause{Lack of Expertise/Education}
\label{sec:lack_of_expertise}
Concerns about a shortage of qualified Solana developers were raised by participants \rnum{C1}, e.g., \participantquote[P6:10]{6}{\ddd In our organization, we are short on Solana developers}. As a result, clients often resort to hiring developers with limited experience, as highlighted by P1: \participantquote[P1:32]{1}{\ddd There's just a lack of developers for all the projects that want something developed. That's why the not-so-experienced developers will also be hired to program something.}
Participant P1 further noted that the focus is set on quick delivery rather than computer science knowledge.
In particular, participants raised concerns regarding the expertise of developers building secure Solana products: \participantquote[P1:32]{1}{Then it's just a lack of knowledge for doing these security checks.} Another participant attributed this issue to a \participantquote[P4:68]{4}{limited education on the Solana security}. They noted (e.g., P5, P2, P4) that educational content on security is needed or that the community needs to be educated about security.

\cause{Lack of Proper Testing/Auditing}
\label{sec:lack_of_auditing}
Thirteen survey participants (e.g., \Ps{7}, \Ps{2}, \Ps{24}, \Ps{31}) reported issues with missing testing and auditing \rnum{C2} (e.g., \simplequote{Poor code review processes; lack of budget or availability for proper audits; Mere Unsafe testing practices like writing unit tests}).
P3 further highlighted: \participantquote[3:7]{3}{Many times, I have made mistakes, and there are a lot of developers like me who have made mistakes due to [a] lack of testing.}

Although code reviews are good practices in industry and open source, five participants (P1, P3, P5, P6, P7) mentioned that audits are not standard in their work as smart contract developers. Participant P7 even described audits as \simplequote{very unusual}. Further, he continued that he only experienced an additional audit once in his career as a developer.
P1 explained that he usually works on small projects but always recommends that his customers do an additional audit. However, he also noted that sometimes the audit would cost more than what could be lost in case of a security breach. Furthermore, P5 mentioned, that customers do not always have a budget for an audit and he also stated in the survey, that he did not check for security mistakes, because we did not ask for a \emph{security} audit. P4 also highlighted, that \participantquote[P4]{4}{that clients don't always allocate a budget for code reviewing}. Participants from our interviews, also those who are part of a larger team, often reached out to external audits (e.g., P2, P4, P6), although some had experienced developers within their own team.


\begin{table}[tb]
\centering
\caption{Security challenges encountered by developers during the development of Solana smart contracts}
\label{tab:vuln_Challenges}
\medskip

\begin{tabular}{@{}llr@{~}l@{}}
\toprule
\textbf{ID} & \textbf{Challenges}    & \multicolumn{2}{c@{}}{\textbf{\# P. Survey}} \\
\midrule
\Cref{sec:lack_of_expertise}     & Lack of expertise/education       & 16 & (\p{45.7}) \\
\Cref{sec:lack_of_auditing}      & Lack of proper testing/auditing   & 12 & (\p{34.3}) \\
\Cref{sec:complexity_rust}       & Complexity of Rust                & 9  & (\p{25.7}) \\
\Cref{sec:young_ecosystem}       & Solana ecosystem is still young   & 8  & (\p{22.9}) \\
\Cref{sec:security_not_priority} & Security is not a priority        & 6  & (\p{17.1}) \\
\Cref{sec:lack_of_resources}     & Lack of supporting resources      & 5  & (\p{14.3}) \\
\bottomrule
\end{tabular}

\end{table}

\cause{Complexity of Rust}
\label{sec:complexity_rust}
Rust stands out for its unique syntax, speed, and functionality not yet available in other languages (e.g., P1, P3, P4). One participant compared Rust to Solidity saying, \participantquote[P3:65]{3}{\ddd many things are not yet there in Solidity but are available in Rust programming languages and pretty fast}.
Its multi-threading capabilities set it apart from other single-threaded languages, compared to JavaScript, with a  participant mentioning, \participantquote[P3:68]{3}{\ddd Rust is not a single-threaded programming language like JavaScript}.
Additionally, Rust's built-in security features, static analysis tools, and memory safety enhanced its safety aspect (e.g., P3, P6, P7). As one participant highlighted, \participantquote[P7:70]{7}{\ddd we have memory safety as one major important thing.}
It provides speed, efficiency, safety and simply being a low-level language made it ideal for decentralized solutions like Solana (P1, P4, P7).
The security features of Rust,
contribute to the prevention of common errors and vulnerabilities (e.g., P2, P6, P7).

During the survey, we sought to examine the influence of Rust's security features on developers' security practices by comparing the time dedicated to security during Rust development to other programming languages.
\pnum{18}{35} reported less time spent on security in Rust compared to another language, while, while \pnum{9}{35} said it's the same and \pnum{8}{35} claimed it takes more time.  
Rust's low-level nature and similarity to C++ were seen as helpful for addressing security concerns (e.g., \Ps{8}, \Ps{26}). Rust's compiler was praised for being a powerful tool that detects security issues before runtime (e.g., \Ps{17}, \Ps{19}, \Ps{4}). It's strong type system and memory safety were listed as key factors in preventing common vulnerabilities like buffer overflows (e.g., \Ps{2}, \Ps{16}, \Ps{17}). 

Despite Rust's advantages, participants highlighted challenges they faced in adopting Rust. The steep learning curve was one of the common concerns, requiring additional time and effort to learn Rust (e.g., P2, P3, P7), with one participant sharing, \participantquote[7:14]{7}{It does take a lot of time to understand as compared to developing different types of software.}. Participants P2 and P5 also noted the lack of comprehensive documentation for Rust, especially in the Solana ecosystem. Further, P2 highlighted the lack of experienced Rust developers which made it difficult to find assistance and guidance. 

To address these challenges, some participants started using development frameworks like Anchor \cite{anchor} to simplify smart contract development and offer support in terms of security (e.g., P1, P3, P6) with a participant claiming that \participantquote[1:14]{1}{anchor makes it a bit easier now}. During the code review tasks, a number of participants (e.g, \Ps{8}, \Ps{10}, \Ps{20}, \Ps{26}) suggested using the Anchor framework for writing Solana smart contracts with one of them stating \surveyquote[P20:5]{20}{Now, most Solana devs are using Anchor framework.
It makes development much easier.}. P1 and P2 mentioned emerging tools such as Neon \cite{neon} and Seahorse Lang \cite{seahorse-lang}, which aimed to facilitate the development of Solana smart contracts using familiar languages like Python.

\cause{Solana Ecosystem is Still Young}
\label{sec:young_ecosystem}
As \rnum{C4}, participants explained that Solana is a rather young community:
\participantquote[P1:32]{1}{\ddd We're more hip. The crowd is also younger. If you go to Solana events, you have the majority of people from 18 to 35.} Another participant shared their experience of joining the community as a student \participantquote[P3:2]{3}{\ddd I was a student looking for an income stream, so at that time, I got to know about blockchains, and I start learning about it.}
As the main reason for joining the Solana community, the quick financial gain was mentioned: \participantquote[P1:60]{1}{\ddd it's attractive for new developers because there's so much money flowing into that space.} Participant P7 described this phenomenon also as a \participantquote[P7:36]{7}{gold rush}.

"Superteams"~\cite{superteamfun} in the Solana community, consisting of committed and long-standing members, play a crucial role in providing support:
\participantquote[P2:10]{2}{There are some small teams that Solana has made in different parts of the world, and they are known as superteams. Those are the community of people who have been in Solana since it started. They are quite active, and they are so helpful that you just have to put up the question over there, put up the smart contract, and they will revise it.} 

Participants also highlighted the importance of engaging with colleagues for code reviews, \participantquote[P4:38]{4}{\ddd one of my immediate approach is consulting a fellow auditor and developer, we get to go through the code}. Further, they demonstrated a sense of responsibility and support for each other's code, one participant mentioning: \participantquote[P1:52]{1}{I also look at my friend's code sometimes or even at the public repos.}

\cause{Security is Not a Priority}
\label{sec:security_not_priority}
Many participants (e.g., P1, P2, P5) also pointed to the fact that security is not the main focus of developers but functionality \rnum{R5}: \participantquote[P2:49]{2}{Like writing the programs, we basically write the contract, and we just focus on one thing, whether it is working or not. Whether we are getting the success message or not, but we rely less on the security part of it.}
Another participant indicated the lack of compensation for security implementation: \participantquote[P1:32]{1}{Even just not caring, I get paid so much, and I make it work. I'm not getting paid to check all the security vulnerabilities because I'm not getting paid more. Why would I do that?}

\cause{Lack of Supporting Resources}
\label{sec:lack_of_resources}
Additionally, five survey participants claimed that there is a lack of security guidelines and training on the security of \ssc, and specifically, Rust C3 (e.g., \Ps{5}, \Ps{14}, \Ps{7}).
Further P5 added, that the documentation of Solana should be improved, as \participantquote[P5:86]{5}{the documentation is really weak at the moment}.

\begin{summary}{RQ 2 -- Summary} 

Developers face challenges in Solana owing to a "gold rush" mentality. The demand for qualified Solana developers is high, and these developers are often young, less experienced, and motivated by quick gains. Unlike "traditional" developers prioritizing security for payment-related products, Solana smart contract developers typically neglect code reviews, testing, and audits unless explicitly requested and paid for by the client. While Rust offers unique security features, its complexity and limited documentation pose challenges, leading developers to adopt new frameworks like Anchor.

\end{summary}

\section{Prevalence of Vulnerabilities in Solana}
\label{sec:tool}
Previous work~\cite{vrust} investigating Solana smart contracts at the source code level cannot be used to estimate the prevalence of vulnerabilities on the Solana blockchain since the number of open-source smart contracts is considerably more limited compared to Ethereum maintaining a database of source code for most of the deployed smart contracts~\cite{etherscan-verified,smart_contract_sanctuary}. In fact, only \pnum{91}{6324} of the \symresDataSize smart contracts on the Solana mainnet reference to source code. 
Hence, studies based on source code only provide a minor glimpse into the actual landscape of vulnerabilities in \ssc. 
As such, performing a study at the bytecode level is essential, allowing analysis of all deployed Solana contracts.
In this work, we focused on the analysis of the prevalence of \acpi, as it was
\begin{enumerate*}[i)]
    \item the most prevalent vulnerability in previous work~\cite{vrust}
    \item the vulnerability where developers rate themselves as the least confident (cf. \cref{sec:code_review_results}), and
    \item only spotted by 2~participants in the code review task of our study (cf.~\Cref{tab:review_results}). 
\end{enumerate*}
In this section, we present our implementation details and evaluate the performance of our framework on currently deployed smart contracts on Solana.

\subsection{Symbolic Execution for Solana}
\label{sec:symexe}
When devising our framework, we opted to rely on symbolic execution because it can overcome the challenges of bytecode analysis.
Symbolic execution is a well-established technique for finding vulnerabilities in computer programs~\cite{Luu2016-dk,Torres2018-je}.
The conditions under which a specific CPI causes a security vulnerability are rather complex, and the combination with low level bytecode makes manual analysis infeasible.
Symbolic execution allows us to gather these constraints automatically to analyze these conditions in detail.
Lastly, smart contracts are reasonably small in code size, so well-known limitations of symbolic execution (e.g., path explosion) remain largely manageable.
To apply symbolic execution to Solana, we load \ssc into a symbolic engine, set up the state according to the Solana input format, and simulate the bytecode until we discover a CPI.
When this exploration reaches a CPI, we evaluate if this CPI is vulnerable.
\begin{figure}[t]
    \centering
    \includegraphics[bb=0 0 255 423, width=.5\linewidth]{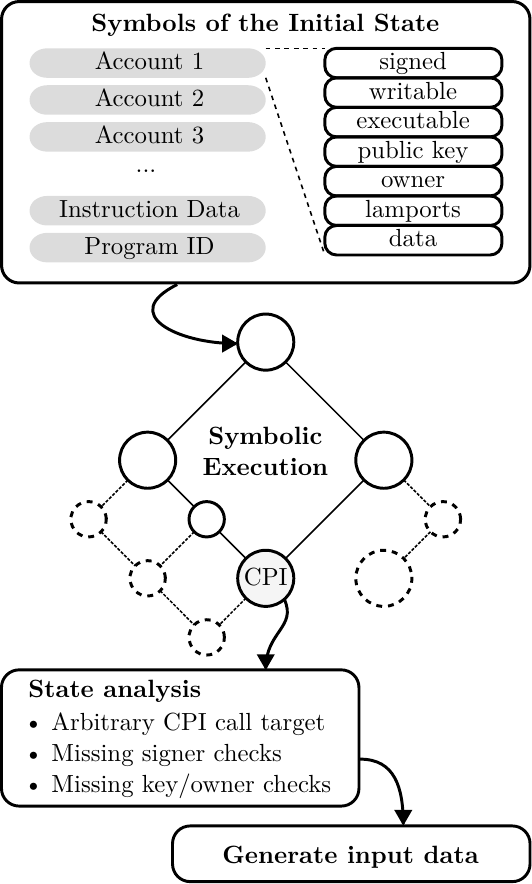}
    \caption{Main concept of the symbolic execution engine.}
    \label{fig:symex}
\end{figure}
\Cref{fig:symex} summarizes the main concept of our framework.
We start with loading the binary into the symbolic execution engine.
Thereafter, we set up the initial symbolic state for the smart contract.

\paragraph{Initial Input State}
As described in \Cref{sec:solana-background}, \ssc do not contain (internal) state.
Thus, contracts can only operate on their input data, allowing an adversary to control every mutable piece of input data to attack a specific contract logic.
This is the reason that contracts have to validate the input data thoroughly.
We set up an initial symbolic state consisting of the symbols shown in \Cref{fig:symex}.
In the following, we describe each symbol in detail:

\begin{description}[format=\normalfont\emph,wide,nosep]
\item[Accounts.] When interacting with \ssc, users specify which accounts should be part of an instruction.
A smart contract can only access accounts that belong to the instruction in which context the contract is called.
Before executing an instruction, the Solana runtime populates the metadata of an account; hence, this information is trusted.
In particular, we use symbolic values for each of the account fields (e.g., signer, owner, and data; see \Cref{fig:symex} and \Cref{sec:SolanaProgramVulnerabilities}).
Note that the runtime only ensures that the account data are correctly loaded from the blockchain.
However, it cannot validate whether the given accounts are also the expected ones by the smart contract.
An attacker can always fabricate accounts and populate them with arbitrary data.
Therefore, it is essential for smart contracts to validate the accounts to mitigate vulnerabilities.

\item[Instruction Data.] This is an \emph{arbitrary} data field used to pass additional parameters to the smart contract.
Typically, this is used to switch between different functionalities of the smart contract.
The Solana runtime does not validate the instruction data; it simply forwards the data given in the transaction to the smart contract.
As a result, an attacker can choose the instruction data arbitrarily.

\item[Program ID.] This is the ID of the called program.
This field is provided by the runtime and can be trusted.
We use a symbol to track when it is used in a comparison, e.g., for an owner check.
\end{description}

\noindent
Starting from this initial symbolic state, we proceed to analyze the target's bytecode.

\paragraph{Symbolic Exploration}
We use an exploration strategy to efficiently reach CPI calls and maximize coverage.
That is, we stop executing paths early when CPI calls are no longer reachable.
These paths are often used for error handling that end in a call to \texttt{abort}.

Therefore, we statically analyze the control-flow graph~(CFG) of the smart contract.
We collect all call sites of the CPI system call provided by the Solana execution environment and automatically infer the control-flow paths that lead to CPI calls during the exploration.
At every state, we test for the following conditions.
First, if the current function contains a CPI call, we guide our engine to this instruction.
Second, if it is possible to reach a CPI call through a callee function, we lead the execution to the basic block that invokes this callee function.
Third, we analyze the stack trace if a function's caller can reach a CPI, directly or indirectly, then we guide the execution the function return.
Otherwise, we abort the execution of this path because no CPI call is reachable from this state.
Whenever our symbolic exploration reaches a CPI call, we proceed with analyzing the state to uncover vulnerabilities.

\paragraph{Vulnerability Analysis (ACPI Oracle)}
We analyze all CPI calls if they are vulnerable.
For a CPI to be an \acpi vulnerability, the call target must be arbitrary, i.e., read from input data, which can originate from any account.
If the contract's logic sanitizes the call target to only allow a few whitelisted targets, then the CPI is not arbitrary.
Otherwise, if the target can assume any address, we consider this an \emph{arbitrary} CPI.
For the \acpi to be a vulnerability, an attacker must be able to control this target/part of the input, which is the case when the contract does not check the trustworthiness of the account in question, either through signer checks or owner checks.

In the case of \emph{Signer checks}, the validity of the check is specific to the smart contract's business logic.
Since we cannot infer the business logic of the contract only from its bytecode alone, we conservatively assume a vulnerability only in cases where all accounts lack a signer check.

In the case of \emph{owner checks}, the contract can ensure whether an account (the address) or its owner is trusted.
Typically, smart contracts will check that the owner of data accounts is the contract itself.
There are three methods to check the owner of an account:
\begin{description}[format=\normalfont\emph,wide,nosep]
\item[Key/Address checks] are often used prior to accessing data that a contract manages only for a specific user, e.g., their stored items or token balance (cf.~\Cref{subsec:methodology_survey}).
\item[Owner field checks] are commonly used to check the authenticity of a contract's data account.
This is particularly important if the data is only read.
\item[Implicit by write.] Writing data to an account is also a valid owner check because the runtime only allows the owner to write data.
\end{description}

Smart contract developers implement both, key and owner checks, through conditional branches, thus, our symbolic execution engine collects both these checks as path constraints.
In contrast, to uncover implicit owner checks, we track whether the contract writes to the data regions of input accounts.
Therefore, we only report contracts as vulnerable to \acpi when all the previous checks fail.
This minimizes the risks of false positives, although the possibility of false negatives is increased.
Note that the developer of a specific smart contract could easily decide from the ACPI reports if there are further owner or signer checks missing with respect to the contract's business logic and also benefit from the remaining analysis.

\paragraph{Implementation}
We use three projects as a starting point for our implementation.
First, an architecture plugin~\cite{bn-ebpf-solana} based on lief~\cite{lief} to lift Solana bytecode into Binary Ninja's intermediate language,
second, Binary Ninja~\cite{binja} itself to calculate control-flow graphs, and
third, SENinja~\cite{seninja}, which is a symbolic execution plugin for Binary Ninja, inspired by angr~\cite{angr} and based on z3~\cite{z3}.
Since the architecture plugin was incomplete and SENinja had no support for eBPF/Solana, we notably extended both projects.
On top of that, we combine these projects into our analysis framework and develop the algorithms for searching critical states, evaluation, and reporting of vulnerabilities as described above.

\subsection{Evaluation}
\label{sec:Evaluation}
We evaluate the effectiveness of our framework on two examples of \acpi, the marketplace contract from our user study (\Cref{sec:StudySmartContract}), and a similar vulnerability in the Neodyme level~4~\cite{neodyme-workshop} smart contract (Appendix~\ref{sec:neodyme-4}). Last but not least, we conduct a large-scale analysis covering all deployed smart contracts in Solana.
\paragraph{The Study's Marketplace Contract}
\bgroup
\newcommand\is{\texttt{invoke\_signed}\xspace}
\newcommand\ci{\texttt{coupon\_info}\xspace}
\newcommand\pp{\texttt{partner\_prog}\xspace}
\newcommand\rc{\texttt{refresh\_credit}\xspace}
Recall that the developer study asked to find the arbitrary cross-program invocation shown in \Cref{subsec:methodology_survey}, specifically in the \rc function in \Cref{fig:marketplace-cpi}.
The symbolic execution shows in its report that the \rc function contains a vulnerable \acpi.
It shows that the call target is the key/address of the second account, and further that the call target is not part of the path constraints.
This means that no comparison between the call target and any trusted information happened.
Therefore, an attacker can choose the target of the call arbitrarily.
Further, the symbolic execution reports that this function contains no signer checks, and while \ci is owner checked, this is not the case for \pp, which is the account used for the CPI.
This vulnerability can be fixed by validating that the \pp is trustworthy, e.g., by verifying that it corresponds to (the trusted) \ci.

\egroup

\paragraph{Large-Scale Analysis}
\label{sec:large-analysis}
To conduct our large-scale analysis, we collect the bytecode of \emph{all} smart contracts that are deployed on the Solana mainnet (as of May 19, 2023).
In total, we analyze \symresDataSize smart contracts. 
The total number of smart contracts deployed on the Solana main chain is notably smaller than the more than 60 million contracts deployed on the Ethereum main chain due to the requirement for smart contract owners in Solana to cover "rent" fees to sustain their contract.\footnote{Owners can also lock sufficient funds in the smart contract to be exempt from rent.} Unpaid rent results in the pruning of the contract.

In our experiment, we aim to provide a first measurement of vulnerabilities in deployed \ssc.
In general, symbolic execution tends to suffer from path explosion that uses exponential memory.
Although we implement countermeasures (i.e., path pruning) into our symbolic execution framework, we still want to provide insights about the broader picture.
Hence, we use a timeout of 30~minutes per contract in this experiment to keep the memory requirements manageable and analyze all \symresDataSize smart contracts on the blockchain.

\begin{figure}[tb]
\centering
\includegraphics[bb=0 0 219 116, width=.67\linewidth]{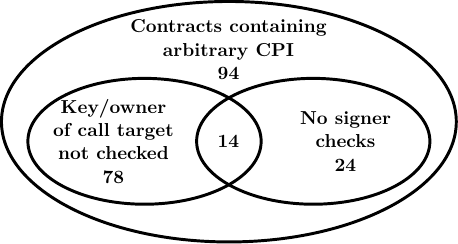}
\caption{Results of the analysis using our symbolic execution framework on deployed \ssc.}
\label{fig:symex-results}
\end{figure}

Due to the lack of source code and accurate decompilation tools, manual verification of the reported vulnerabilities is extremely challenging. 
Nevertheless, we point out that we are highly conservative when deeming that a smart contract is vulnerable. As \Cref{fig:symex-results} shows, the 14~cases that remain in the intersection are highly likely to be genuine positives, considering the inherent risk associated with performing an \acpi without any form of verification on the target contract. We are, however, unable to assess the false negative rate given the absence of a ground truth dataset.\footnote{Since the vulnerable \ssc were deployed fully anonymously without any contact information, it was not possible to contact the respective owners to report these vulnerabilities---even for the maintainers of Solana.
Therefore, we refrained from reporting the IDs of the vulnerable contracts.} That said, our results corroborate previous findings~\cite{smolka2023fuzz} that also witnessed a small prevalence of vulnerabilities in Solana smart contracts.

\begin{summary}{RQ 3 -- Summary}
In total, our symbolic execution framework found arbitrary CPI calls in \symresACPI smart contracts.
\Cref{fig:symex-results} shows the composition:
\symresOwner~of these contracts do not perform key or owner checks, and in \symresSigner cases, contracts do not perform any signer check.
As we argue in \Cref{sec:symexe}, either check may validate trust in the called program, depending on the business logic.
Hence, we consider \symresVuln~cases in the intersection vulnerable because these smart contracts do not validate the call target in any way.That said, the prevalence of arbitrary CPI vulnerabilities in Solana seems to be limited to 0.2\% of all smart contracts in Solana.
\end{summary}

\section{The Increasing Reliance on Anchor}
\label{sec:Anchor_Reliance}
We believe that developers use frameworks, such as Anchor~\cite{anchor}, that help bridge their lack of security experience. This was indeed mentioned by some of the participants in our user study.

To this end, we analyzed all contracts in Solana up to October 29 2023, with the aim to distinguish contracts compiled with the Anchor framework and evaluated all the contracts deployed (and still available) of Solana. We were able to distinguish the bytecode of contracts compiled with Anchor by looking for Anchor-specific error strings and ABI within the contracts; these differ drastically in Anchor from those generated in native Rust.
Our analysis revealed that Anchor has become the defacto standard framework for devising smart contracts in Solana, accounting for 88\% of the deployed contracts at the time of writing.

Given the ever-increasing reliance on Anchor, we extended our analysis to evaluate the prevalence of Anchor usage among the vulnerable contracts identified in \Cref{sec:large-analysis}. Our results show that, indeed, \emph{none of the 14 identified vulnerable contracts were compiled with the Anchor framework}.
Similarly, all 92 vulnerabilities discovered by \citewauthor{smolka2023fuzz} pertained to smart contracts that were not compiled with Anchor.
These findings lend strong support that Anchor has a significant impact on the security of Solana.

\begin{figure}[tb]
\centering
\includegraphics[bb=0 0 406 293, width=.77\linewidth]{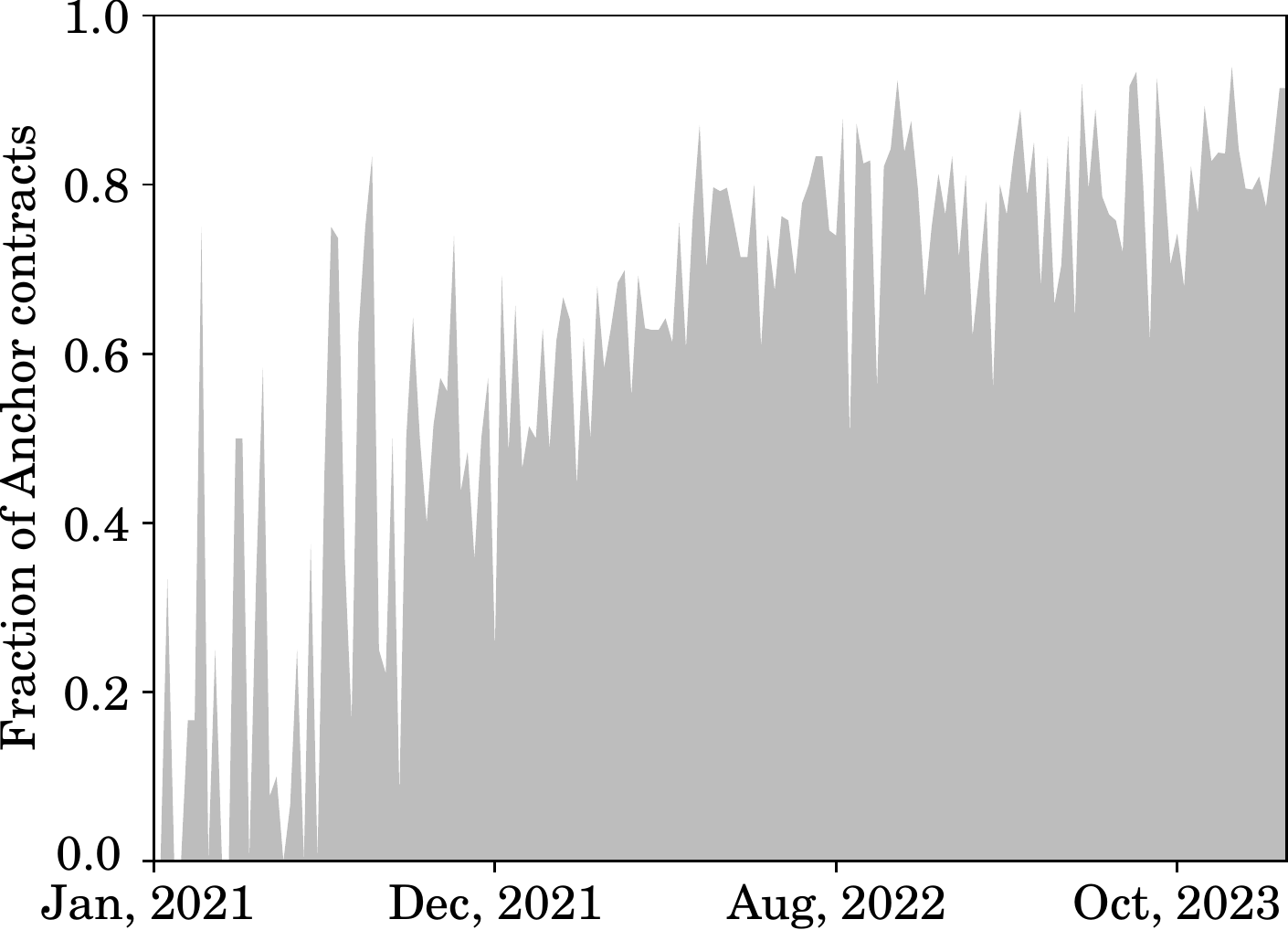}
\caption{Prevalence of Anchor contracts in all the deployed contracts \ssc.}
\label{fig:anchor-usage}
\end{figure}


\section{Related Work}

\paragraph{Security Studies with Developers}
Security vulnerabilities in software are a major issue~\cite{baca2009static}, highlighting that developers are rarely security experts, often lacking usable resources, tools, and a security culture~\cite{GreenS16, AcarS16, acar2017developers, witschey2014technical}. They face challenges in dealing with security in different stages of software development, from designing, coding, and testing to maintenance~\cite{wurster2008developer, xie2011programmers, Assal2018SecurityIT}. 
In their study on the reasons behind developers causing security vulnerabilities, \citewauthor{oliveira2014s} found that security vulnerabilities are considered "blind spots" focusing rather on functionality~\cite{naiakshina2017developers,gutfleisch22}.
While prior studies identified code reviewing as helpful to detect security vulnerabilities~\cite{bosu2014identifying}, 
in a code review study by \citewauthor{danilova2021code}, it was found that only \p{30} of their participants detected security vulnerabilities. 

To the best of our knowledge, this study is the first to investigate security challenges in Solana smart contract development with Rust. Prior research by \citewauthor{parizi2018smart} and \citewauthor{sharma2022exploring} focused on Solidity, Pact, Liquidity, and security practices in Ethereum. While Solidity was considered most usable, participants often made security errors. \citewauthor{sharma2022exploring} interviewed 29 participants about security practices and observed a 25-minute code review for a vulnerable smart contract.
Here, participants were, however, prompted about security, potentially creating a priming effect that could have influenced code reviews and biased the results. 
In our study, we opted not to prime the participants, to better capture real-life deployments.

\paragraph{Smart Contract Security Analysis}
Smart contract security research has produced many approaches to detecting bugs in smart contracts~\cite{Schneidewind2020-mj, harvey, confuzzius, evmpatch, sereum, Mossberg2019-xp, rodler2023efcf} based on multiple techniques, such as formal verification, runtime detection and fuzzing.

Symbolic execution is a common technique for finding critical vulnerabilities in software, that has also been applied to Ethereum smart contracts~\cite{Luu2016-dk,Kalra2018-mq,Torres2018-je,Mossberg2019-xp}, targeting vulnerabilities such as transaction-ordering dependencies, \iobug, and reentrancy.
Although these approaches are successful, they focus on bugs that are not present in Solana, or \iobug that can be trivially mitigated. Last, they are specifically geared for Ethereum. 

Recently, a static analysis tool for \ssc has been proposed.
VRust~\cite{vrust} detects common vulnerability patterns in the source code of \ssc, like MSC, IB, and ACPI.
Although the VRust's source-based analysis could easily fit into development workflows, it is unlikely to be beneficial in practice, as its evaluation shows that VRust suffers from a high false alarm rate of \p{89}.
This overwhelming number of false positives means developers are likely to miss real bugs or ignore the alerts altogether.
Further, VRust cannot be used to facilitate large-scale research on the Solana platform because it relies on source code, which is unavailable for the vast majority (\p[6324][1]{6233}) of \ssc.
Recently, a coverage-based fuzzer, dubbed FuzzDelSol~\cite{smolka2023fuzz}, for Solana programs was able to uncover 92 bugs in 52 programs (roughly 0.8\% of all 6049 considered smart contracts).

\paragraph{Rust}
Rust~\cite{rustlang}, the programming language for Solana smart contracts, was ranked as the most loved language by \p{87} of developers according to the 2022 Stack Overflow survey~\cite{stackoverflowsurvey2022}. Rust is known for its memory and concurrency safety features, achieving performance similar to C/C++. Memory safety is crucial, contributing to nearly \p{70} of security bugs in large projects such as Chromium~\cite{chromium} and Microsoft internal projects~\cite{microsoft}.

Several user studies investigated the challenges posed by Rust for newcomers. \citewauthor{zhu2022learning}'s mixed-methods study revealed developers' difficulties with Rust's concepts. Understanding error messages proved challenging, potentially impeding the learning process. However, \citewauthor{fulton2021benefits} discovered a distinct outcome, suggesting that developers gained confidence from Rust's compiler, which is effective at identifying potential bugs. The study also highlighted increased developer confidence in Rust, as they firmly believe that successful compilation ensures code safety and correctness. \citewauthor{ferdowsi2023usability} examined Rust's usability, identifying challenges for developers with its complex syntax, advanced type system, ownership, and lifetime rules.

\section{Discussion}

\noindent \textbf{Code reviews.}
Despite code reviews being one of the most effective and commonly used techniques among smart contract developers~\cite{Zou, Chakraborty}, only 20\% of our participants were able to identify at least one security vulnerability during their code reviews. 
We found only a weak negative correlation between our participants' experience reviewing smart contracts and their ability to detect security vulnerabilities (see Section~\ref{sec:code_review_results}). This finding is consistent with research indicating that security vulnerabilities often persist even with experienced code reviewers~\cite{Paul, Bosu}. Previous studies on code reviews have shown that security is often not the primary focus. Instead, developers tend to prioritize functionality over security despite the security-critical nature of the task~\cite{danilova2021code}. Our results are also consistent with these findings, showing that participants in our study did not prioritize security during code review tasks, despite the financial nature of the \ssc and the need to protect them against attacks.

\vspace{0.5 em}\noindent \textbf{Solana vs. Ethereum.}
Early research on the prevalence of vulnerabilities in Ethereum smart contracts~\cite{Kalra2018-mq, Luu2016-dk} reported a vulnerability rate of 78\% in a dataset of 1524 deployed contracts.
More recent larger-scale studies with broader datasets consisting of 24595~\cite{Tsankov2018-xu} and 141k~\cite{Brent2018-sl} smart contracts deployed on the Ethereum platform report vulnerability rates of 67\% and 50\% for single vulnerability types.
This sharply contrasts with the 0.2\% prevalence rate that we detected in Section~\ref{sec:large-analysis}. Similarly, another recent study on Solana contracts~\cite{smolka2023fuzz} reported that only 0.8\% of Solana projects were vulnerable. 
\citewauthor{sharma2022exploring} conducted a similar code review study with Ethereum developers. Their study also yielded slightly better results, with 55\% of the interviewees and 20.5\% of survey participants identifying vulnerabilities. 
However, both developer communities shared similar challenges, such as the lack of documentation, shortage of qualified developers, and budget restrictions. 
First, while Ethereum provides more documentation than Solana, developers from both communities expressed a desire for more comprehensive documentation~\cite{sharma2022exploring}. It is worth noting that \citewauthor{Acar2016} found that official documentation is often not used, leading developers to favor quicker alternatives like Stack Overflow, possibly compromising security. 
Second, the scarcity of specialists might prompt smart contract owners to hire unqualified developers, increasing the risk of developing insecure code. Further, due to the shortage of qualified developers, conducting thorough code reviews before deployment may be limited, affecting security.
Third, the perceived prioritization of functionality over security might also influence the submissions. Security is considered as \enquote{something extra,} usually requiring to be requested explicitly by the client and compensated for accordingly (see Section~\ref{sec:security_not_priority}). 
This might also explain the slightly better detection rates from \citewauthor{sharma2022exploring}, as they specifically prompted participants on security. 
Both developer communities highlighted neglecting security considerations regarding budget allocation and prioritization despite the critical nature of smart contracts in handling financial transactions~\cite{sharma2022exploring}.
Past research on software development found similar challenges, such as neglecting security in terms of budget allocation~\cite{assal2019think, Oliveira, Lopez} and a lack of education and expertise~\cite{assal2019think, weir_apps_2020, Assal2018SecurityIT}. However, the shortage of qualified developers in the Solana community seems to stem from both a lack of expertise in the Rust programming language and security (see Sections~\ref{sec:complexity_rust} and \ref{sec:lack_of_expertise}).

\vspace{0.5 em}\noindent \textbf{Rust \& Vulnerabilities.}
To further set these results in context, prior work~\cite{Paul} classified what type of vulnerability is more likely to be discovered during code reviews according to their CWE~\cite{cwe} type. We matched our vulnerabilities \iobug and \msc to their closest CWE match, respectively CWE-682 Incorrect Calculation and CWE-284 Improper Access Control. There was no direct match for \acpi. However, since it is a vulnerability that mostly arises from a lack of verification of the authenticity of the input address, we compared it to CWE-345 (Insufficient Verification of Data Authenticity). 
The detection rate by participants in our study of at least one of the \iobug, \msc, and \acpi vulnerabilities were respectively 11.5\%, 11.4\%, and 5.7\%. On the other hand, the corresponding CWE detection rate in~\cite{Paul} was 79.27\%, 12.12\%, and 0\%. While the latter 2 vulnerabilities were detected with similar accuracy, we could see a significant discrepancy in the detection rate of \iobug. We believe that this might arise (1) from the perception that Rust helps prevent vulnerabilities like buffer overflows, as mentioned by participants, and (2) Rust's inconsistent behavior regarding integer overflows and underflows, which trigger errors in development mode but not in release mode.
Our participants praised Rust for its speed and security features, resulting in increased confidence in their security approach, consistent with the findings of~\citewauthor{fulton2021benefits}. In contrast, Ethereum smart contract developers expressed less satisfaction with Solidity’s security features, highlighting its limitations~\cite{sharma2022exploring}.

\vspace{0.5 em}\noindent \textbf{Anchor.}
Following the relatively safe state of the Solana ecosystem (see \Cref{sec:Evaluation}) in spite of the security challenges faced by developers (see \Cref{sec:Security_Challenges}), we investigated the usage of Anchor and noticed its widespread use, as up to 88\% of \ssc deployed made used of Anchor (see \Cref{sec:Anchor_Reliance}). Our findings suggested that Anchor might be one of the main factors shielding Solana from vulnerabilities introduced by developers who lack expertise in Rust and security. Indeed, none of the vulnerable smart contracts detected by our symbolic execution tool (nor by~\cite{smolka2023fuzz}) were compiled with Anchor. It is plausible that the increased reliance on Anchor has prevented developers from learning Rust in detail and might have impacted the performance in the code review tasks.

\subsection*{Future Research \& Recommendations }
Reflecting on our findings and the challenges we faced during our study, we propose several recommendations.

\vspace{0.5 em}\noindent \textbf{Investigating Anchor.} Anchor's popularity within the \ssc developers community suggests that it plays a vital role in simplifying \ssc development. Although Anchor offers security benefits, it does not protect against all attacks~\cite{anchor-docs}. Thus, it is essential to investigate the security and usability of Anchor, encompassing as well the main reasons behind its widespread adoption. Future studies should investigate the precise role that Anchor plays in affecting the work of \ssc developers and their perception of security. Additionally, it is crucial for Anchor to recognize its vital role within the Solana community.

\vspace{0.5 em}\noindent \textbf{Security Prompting.} 
Our participants advised \ssc owners to prioritize security by specifically including security requirements and allocating a specific budget for professional security audits rather than treating them as ad-hoc activities. 
This recommendation aligns with past research~\cite{naiakshina2017developers}, suggesting that proactive measures influence security outcomes.

\vspace{0.5 em}\noindent \textbf{Recruitment Challenges.}
We encountered significant difficulties in our recruitment process, resulting in a high number of spammers and bots. Similar challenges were faced by \citewauthor{sharma2022exploring} in their Ethereum study. One effective approach to recruiting qualified developers, as demonstrated by \citewauthor{danilova2021really}, is the use of screening questions for evaluating the programming skills of participants. 
Further research is needed to assess new recruitment approaches for blockchain developers. In addition to target group-specific screeners, one potential approach could involve closer collaboration with core communities and their active involvement in research projects.


\section{Conclusion}
In this work, we conducted a study with 35 Solana developers to explore how Solana smart contract developers handle security and the challenges they encounter. Our study revealed a general lack of awareness regarding security vulnerabilities, with almost 83\% of participants approving vulnerable code for release. 
To view the effect of these challenges on the overall security of the ecosystem, we further conducted an automatic analysis of currently deployed Solana smart contracts.
Our findings show a low prevalence of security vulnerabilities in Solana, with only around 0.2\% of deployed smart contracts found to be vulnerable. We also measured an increasing reliance on Anchor, with almost 88\% of \ssc written using it, suggesting that Anchor might play a significant role in safeguarding Solana. 
Therefore, we hope that our work informs new research areas and practices in this ecosystem, especially focusing on Anchor's contribution to preserving an ecosystem from common security vulnerabilities.
Finally, we call for 
further research with developers working on other blockchain platforms to comprehensively understand smart contract development security practices in other ecosystems beyond Solana and Ethereum.

\section*{Acknowledgement}
The authors would like to thank the anonymous reviewers for their valuable comments and suggestions. We are additionally grateful to Salih Kardag for investigating the usage of the Anchor framework.
This work has been partially funded by the Deutsche Forschungsgemeinschaft (DFG, German Research Foundation)---EXC 2092 (CASA) 39078197, SFB 1119 (CROSSING) 236615297 within project T1---and the European Union through the HORIZON-JU-SNS-2022 NANCY project with Grant Agreement number 101096456 and the ERC CONSEC project, Grant Agreement number 101042266. Views and opinions expressed are, however, those of the author(s) only and do not necessarily reflect those of the European Union or the SNS JU. Neither the European Union nor the granting authority can be held responsible for them.

\printbibliography

\appendix

\numberwithin{figure}{section}
\numberwithin{table}{section}




\section{Replication Package}
\label{sec:replication}
To make our study reproducible, and allow for easy access for meta research, we publish a replication package containing the following documents:
\begin{enumerate}[(1)]
\item The explanation to prepare for the review task, as well as the task description for each file of the smart contracts (\ref{sec:appedix_review_task});
\item The post-task questionnaire \ref{sec:appendix_posttask};
\item The interview guide with the main questions and follow-up prompts for the semi-structured interviews (\ref{sec:appendix_interview_guide});
\item The operationalized code book used for coding the interviews with codes contributing to this paper (\Cref{tab:codebook});
\item The source code of both smart contracts used in our study: \url{https://figshare.com/s/216e99bf71e62db7889f}\,.
\end{enumerate}

\titleformat*{\subsection}{\normalsize\bfseries}


\subsection{Code Review Task}
\label{sec:appedix_review_task}

\subsubsection{Pre-Task Instructions}
For the upcoming \textbf{three tasks} you should \textbf{conduct a code review} for the code snippets provided.
The code snippets provided belong to one Solana smart contract. 
Please review them as you usually would when reviewing code for Solana smart contracts. You can review the code manually or use any tools you need to conduct your usual code review. You get to decide if this code is eligible for release or if it needs to undergo certain improvements to be eligible for release.
\\
The following 3 code snippets belong to one Solana smart contract. The analysis target is a trading subsystem that implements marketplace logic for some Solana app. The marketplace logic follows a microtransaction pattern, so a user would buy an in-app currency (which we just call tokens) to spend it on items. The project is organized into 5 files. The `lib.rs` is the umbrella for the project, i.e., it does nothing on its own and can be safely ignored. The `entry.rs` file defines the entry point logic, i.e. instruction parsing and dispatching, and also provides basic user management.
These code snippets provided can be:
\begin{itemize}
    \item Copied as text
    \item Copied from Gitlab
    \item Viewed or downloaded from the survey
\end{itemize}
Make sure your \textbf{work set up is ready} to conduct a proper code review before clicking on next.
Please conduct the whole survey at once without interruptions.

\noindent[Newpage]

\subsubsection{Review Task Instructions\protect\footnote{Each file (tokens.rs, items.rs or partners.rs) is presented at a separate page. This also includes file related explanations.}}

\noindent\warning \space  Please conduct this task as you would usually do when reviewing code for Solana smart contracts. 

\noindent\warning   \space  Please only do a code review for the file [tokens.rs, items.rs, partners.rs].\\

\noindent[tokens.rs] The `token.rs` file provides `TokenVault` that manages the amount of tokens that a user owns, i.e., it provides functionality to create vaults (on behalf of a user), to buy, sell, transfer and distribute (i.e., transfer the same amount of tokens to multiple users at once) tokens.\\

\noindent[items.rs] The `item.rs` file provides `Item` and `Offer` structures to user for trading items. It provides functionality to create offers and to accept offers for items -- with offers coming from the buyer.\\

\noindent[partners.rs] The `partner.rs` file provides some voucher/coupon-like logic, i.e. you can think of this as partners airdropping tokens to the app running this marketplace.\\

\noindent The file can be found \underline{here} on GitLab.\\
\noindent The full project can be cloned \underline{here} on GitLab.\\

\noindent[Embedded Gitlab snipped: tokens.rs, items.rs, partners.rs]\\

\noindent[Expandable table with free text fields, in which the \emph{line number}, \emph{the comment} and \emph{the fix} can be entered.]\\

\paragraph{Q0} Would you approve this code for release in its current state?
\begin{enumerate*}
    \renewcommand{\labelitemi}{\scriptsize$\bigcirc$}
    \item Yes, the code is releasable without any risks.
    \item Yes, the code is releasable minor risks.
    \item No , the code needs improvements before releasing.
\end{enumerate*}

\paragraph{Q0.1} Please explain your decision [Free text field]\\

\noindent[Newpage]


\subsection{Post-Task Questionaire}
\label{sec:appendix_posttask}
{\small
Thank you for reviewing the smart contract. You will proceed with the questionnaire now.\\

\noindent Please, from now on, refrain from using google or any other search engines \/ other information sources. Please try to answer the questions to the best of your knowledge.

\paragraph{Q-Attention} 
This is an attention check question. Please select the answer ”Octal”.
\begin{enumerate*}
    \renewcommand{\labelitemi}{\scriptsize$\bigcirc$}
    \item Duodecimal
    \item I don’t know
    \item Octal
    \item Binary
    \item Decimal
    \item Hexadecimal
\end{enumerate*}

\noindent[Newpage]

\paragraph{Q1} Which option was used to conduct the code review?
\begin{enumerate*}
    \renewcommand{\labelitemi}{\scriptsize$\bigcirc$}
    \item Copied text
    \item Cloned files from GitLab
    \item viewed or downloaded from the survey
    \item Other
\end{enumerate*}

\paragraph{Q2} How much time did it take to set up the appropriate environment to fulfill each of the tasks task? (please fill in the time in minutes)

\paragraph{Q3}How much time on average did each task take? (please fill in the time in minutes)

\paragraph{Q4} Did you use any guideline while performing the code review for the three tasks?
\begin{enumerate*}
    \renewcommand{\labelitemi}{\scriptsize$\bigcirc$}
    \item Yes
    \item No
\end{enumerate*}

\paragraph{Q4.1} What kind of guideline(s) did you use? (multiple selections are possible)
\begin{enumerate*}
    \renewcommand{\labelitemi}{\scriptsize$\bigcirc$}
    \item Internal company Guideline
    \item Personal guideline
    \item Online Guideline \underline{\hspace{1cm}}
    \item Other Guideline  \underline{\hspace{1cm}}
\end{enumerate*}

\paragraph{Q5}How did you review the code snippet for the three tasks? (multiple selections are possible)
\begin{enumerate*}
    \renewcommand{\labelitemi}{\scriptsize$\bigcirc$}
    \item Manually
    \item Using the following tools \underline{\hspace{1cm}}
\end{enumerate*}

\paragraph{Q6} What was your primary focus while conducting the code review for all three tasks?

\paragraph{Q7} Did you specifically check for security vulnerabilities while performing a code review for the three tasks?
\begin{enumerate*}
    \renewcommand{\labelitemi}{\scriptsize$\bigcirc$}
    \item Yes. What did you check for? \underline{\hspace{1cm}}
    \item No
\end{enumerate*}

\paragraph{Q8}In a real scenario would you have asked for advice on the smart contract (e.g. from a colleague or friend)?
\begin{enumerate*}
    \renewcommand{\labelitemi}{\scriptsize$\bigcirc$}
    \item Yes
    \item No
\end{enumerate*}


\noindent[Newpage]

\paragraph{Q9} How many times have you reviewed Solana smart contract code written by others so far? 
\begin{enumerate*}
    \renewcommand{\labelitemi}{\scriptsize$\bigcirc$}
    \item More than [Sliding scale from 1 to 100]
\end{enumerate*}

\paragraph{Q10} What percentage of your \textbf{\underline{code reviewing time}} do you usually dedicate to security?
\begin{enumerate*}
    \renewcommand{\labelitemi}{\scriptsize$\bigcirc$}
    \item{} [Sliding scale from \p1 to \p{100}]
\end{enumerate*}

\paragraph{Q11} What percentage of your \textbf{\underline{programming time}} do you usually dedicate to security?
\begin{enumerate*}
    \renewcommand{\labelitemi}{\scriptsize$\bigcirc$}
    \item{} [Sliding scale from \p1 to \p{100}]
\end{enumerate*}

\paragraph{Q12} How much time do you spend on the code's security when developing with Rust compared to other programming languages?
\begin{enumerate*}
    \renewcommand{\labelitemi}{\scriptsize$\bigcirc$}
    \item Less time.  Please explain your choice: \underline{\hspace{1cm}}
    \item Equal time.  Please explain your choice: \underline{\hspace{1cm}}
    \item More time. Please explain your choice: \underline{\hspace{1cm}}
\end{enumerate*}

\paragraph{Q13} Please carefully read each statement and state how much you agree with the respective statement. [7-Point scale, 1= Strongly disagree, 7= Strongly agree] 
\begin{enumerate*}
    \renewcommand{\labelitemi}{\scriptsize$\bigcirc$}
    \item I have a good understanding of security concepts in Smart Contracts
    \item I have a good understanding of security concepts in Rust
    \item I feel responsible for the security of end-users when writing code      
    \item I feel responsible for the security of end-users when reviewing code
\end{enumerate*}

\noindent[Newpage]


\paragraph{Q14} Please rate how confident you are in your ability to explain the following Solana smart contract security vulnerabilities and its mitigation strategies to a colleague. [5-Point scale, 1= not confident at all, 5= completely confident] 
\begin{enumerate*}
    \renewcommand{\labelitemi}{\scriptsize$\bigcirc$}
    \item Missing ownership check
    \item Missing signer check
    \item Integer overflow and underflow    
    \item Arbitrary signed program invocation
    \item Solana account confusions
\end{enumerate*}

\paragraph{Q15} Did you check for one of the above security vulnerabilities in the three code review tasks?
\begin{enumerate*}
    \renewcommand{\labelitemi}{\scriptsize$\bigcirc$}
    \item Yes. Can you please list which one(s): \underline{\hspace{1cm}}
    \item No. Can you explain why? \underline{\hspace{1cm}}
\end{enumerate*}

\noindent[Newpage]

\paragraph{Q16} The previously mentioned security vulnerabilities were spotted in deployed Solana smart contracts. What do you think are the reasons that these vulnerabilities are so prominent in many Solana smart contracts? [free text field]
\noindent[Newpage]

\subsubsection{Demographics}

\paragraph{Q17} How old are you?

\paragraph{Q18} What is your gender?
\begin{enumerate*}
    \renewcommand{\labelitemi}{\scriptsize$\bigcirc$}
    \item Female
    \item Male
    \item Non-binary/ third gender
    \item Prefer not to answer
    \item Prefer to self-describe:  \underline{\hspace{1cm}}
\end{enumerate*}

\paragraph{Q19} What is the highest level of school you have completed or the highest degree you have received?
\begin{enumerate*}
    \renewcommand{\labelitemi}{\scriptsize$\bigcirc$}
    \item Less than high school / GCSE or equivalent
    \item High school or equivalent / A level or equivalent
    \item Vocational degree
    \item Bachelor's degree
    \item Some graduate school, or currently enrolled in graduate school
    \item Master's or professional degree
    \item Doctorate degree
    \item Other  \underline{\hspace{1cm}}
\end{enumerate*}

\paragraph{Q20} In which country do you reside in? [Free text field]

\paragraph{Q21} What is your current employment status?
    \begin{enumerate*}
    \renewcommand{\labelitemi}{\scriptsize$\bigcirc$}
    \item Employed full-time
    \item Employed part-time
    \item Full-time student 
    \item Retired
    \item Other (please describe.)  \underline{\hspace{1cm}}
    \item Prefer not to answer
\end{enumerate*}

\paragraph{Q22} What is your current job title?

\paragraph{Q23} How many years have you worked in the software industry?

\paragraph{Q24} How many years have you worked as a software developer?

\paragraph{Q25} How many years of experience do you have developing Solana smart contracts?
    
\paragraph{Q26} How many years of experience do you have reviewing Solana smart contracts?
\paragraph{Q27} How many years of programming experience do you have in Rust?
\noindent[Newpage]

\subsection{Off-boarding Questions}

\paragraph{Q28}We are interested in offering fair compensation for your participation in our research. How do you rate the payment (50\$) of the study? 
\begin{enumerate*}
    \renewcommand{\labelitemi}{\scriptsize$\bigcirc$}
    \item Way too little
    \item Too little
    \item Just right
    \item Too much
    \item Way too much
\end{enumerate*}

\paragraph{Q29} How many minutes did you actively work on this survey? (This will not impact the amount of your compensation)

\paragraph{Q30} We are interested in learning more about the development of smart contracts by \textbf{conducting a 45-minute online interview via Zoom.} Participation will be compensated \textbf{with an additional 50~USD (50\,€)} (Amazon voucher). If you are interested, please fill in your email address in order to contact you in case you were selected. (You are not obliged to conduct the interview in case you changed your mind)
\begin{itemize}
    \renewcommand{\labelitemi}{\scriptsize$\bigcirc$}
    \item I am interested, here is my email \underline{\hspace{1cm}}
    \item I am not interested in conducting an interview
\end{itemize}

 \paragraph{Q31} Please chose the method of payment for your compensation and provide the adequate email address for receiving it. 
 \begin{enumerate*}
    \renewcommand{\labelitemi}{\scriptsize$\bigcirc$}
    \item Amazon voucher\underline{\hspace{1cm}}
    \item Paypal \underline{\hspace{1cm}} 
    \item I don't want to receive a compensation
    \item I will contact you before [x] via email
\end{enumerate*}

\paragraph{Q33} If you have any further questions or comments regarding the study please let us know. \underline{\hspace{1cm}}\\
 
\noindent[Newpage]

\noindent We would like to explain to you the goal of the study:
Our study was motivated by vulnerabilities spotted in Solana smart contracts. With this work, we wanted to investigate these vulnerabilities further and explore the abiltiy to detect them before releasing the code.

\noindent Thanks for participitating in our study. 

}


\begin{table*}[tbp]
\caption{Our final codebook. (*) denotes a container for sub-codes and therefore is not used during coding.}
\label{tab:codebook}
\centering
\scriptsize
\renewcommand{\arraystretch}{1.1}
\setlength{\tabcolsep}{0.6\tabcolsep}
\setlength{\defaultaddspace}{0.33\defaultaddspace} 
\rowcolors{2}{white}{gray!10}
\scalebox{0.9}{\begin{tabularx}{\linewidth}{lp{.35\linewidth}>{\raggedright\itshape\arraybackslash}X}
    \toprule
    \textbf{Code} & \textbf{Description} & \textbf{\textup{Example Quote}} \\
    \midrule
    \textbf{Participant Background} & Statements that describe the professional background (education, work experience…) of the participant. & \codebookquote[(Interview 18: 15)]{18}{"I'm an experienced blockchain developer, also an expert both in backend and front end technology, so you can call me a full stack blockchain developer"} \\
    \textbf{Software Development Process} & General statements made by the participants regarding their software development process (from design to maintenance). & \codebookquote[(Interview 23: 66)]{23}{"It's basically making an architecture first. What things should go on the smart contract and how it will be implemented, and then decide whether it will be accessible by just the owner, or it will be a multi-sig contract."} \\
    \textbf{SDP Security Measures } & Statements that describe the security measures that the participants take while developing smart contracts. & \codebookquote[(Interview 23: 66)]{23}{"Every time we write a method, we ensure like there are several security checks that we need to ensure like signer is authorized."} \\
    \textbf{Solana Development Comparison} & Statement that compares Solana to other Blockchains or Software & \codebookquote[(Interview 23: 66)]{23}{"The biggest difference is I would like to draw this comparison to something like standard smart contracts that are based in JS. What I really think is that it's not really well documented."} \\
    \textbf{Reviews and Audits} & General statements made by the participants regarding audits and reviews including peer reviews, the benefits of a review, the need for an audit… & \codebookquote[(Interview 23: 66)]{23}{"The things that I build don't involve much value. It's usually just an NFT modification smart contract or something like that where the value that we handle is much lower than an audit would cost"} \\
    \textbf{Rust*} &  &  \\
    \qquad\textbf{Difference to other languages} & Rust comparison to other languages. & \codebookquote[Interview 13: 83]{13}{"Well, obviously, the speed or the closeness to hardware. We have similar low-level possibilities like we have in C and it feels like a higher language than C."} \\
    \qquad\textbf{Challenges} & Challenges faced when adopting Rust as a programing language. & \codebookquote[Interview 9: 57]{9}{"The syntax is quite new for everyone. They don't see similar syntax anywhere in other languages."]}\\
    \qquad\textbf{Advantages and (Security) Features} & Advantages and features gained when adopting Rust as a programing language including its security features. & \codebookquote[Interview 21: 194]{21}{"I like the compiler a lot because it tells me where the error is and also what I should do to get rid of that error."}\\

    \textbf{Improvements} & Participants; suggestion for improving the development of secure Solana smart contracts & \codebookquote[(Interview 23: 66)]{23}{"I guess a lot of education has to be done. Teaching people the basics of security or even just the basics of how smart contracts work on Solana or programs work on Solana and what people need to look out for."} \\
    \textbf{Advice for securing smart contracts} & Participant's advice to developers to ensure the development of secure Solana smart contracts & \codebookquote[(Interview 23: 66)]{23}{"Know what you're building so that you can know what to protect"} \\
    \textbf{Security Challenges} & General statement regarding the security challenges \ssc developers face in the Solana ecosystem & \codebookquote[(Interview 23: 66)]{23}{"Rust is quite hard, and people don't want to go the long way to learn it and develop it. They just use some shortcuts."} \\
    \textbf{Security Vulnerability Knowledge} & Participants description of some security vulnerabilities and their mitigation, including Arbitrary assigned program invocation, Integer Overflow/Underflow and Missing signer check. & \codebookquote[(Interview 23: 66)]{23}{"if you want some action to only be allowed by a certain authority, then you need to make that authority a signer."} \\
    \textbf{SC development frameworks} & General information regarding other tools that support in the development of Solana smart contracts & \codebookquote[(Interview 23: 66)]{23}{"This anchor gives you the high level capabilities to write all the things in short and sweet manner where you can differentiate between your validation part and the business logic part."} \\
    \textbf{Solana community} & General information about or related to the Solana community  (e.g.. developer's characteristics, type of support offered in the community, ...) & \codebookquote[(Interview 23: 66)]{23}{"Basically, it's the community feedback. There are some small teams that Solana has made in different parts of the world, and they are known as Superteams. Those are the community of people who have been in Solana since it started."} \\
    \textbf{Solana Characteristics} & General Information regarding the characteristics of the Solana Blockchain & \codebookquote[(Interview 23: 66)]{23}{"they chose scalability and security."} \\

    \bottomrule
\end{tabularx}}
\end{table*}


\begin{table*}[!htb]
	\caption{Overview over interview participants. ($^\dagger$ ISO~3166-1 encoded.)}
	\label{tab:participants}
	\centering
	\footnotesize
	\setlength{\tabcolsep}{2pt}
	\rowcolors{2}{white}{gray!10}
\scalebox{0.9}{\begin{tabular}{@{}lllllccc@{}}
\toprule
   & \textbf{Highest Degree} & \textbf{Country$^\dagger$} & \textbf{Current Job Status} & \textbf{Current Job Status} & \textbf{Software Dev. Exp.} & \textbf{Solana Exp.} & \textbf{Rust Exp.} \\ \midrule
P1 & Master                  & AT               & Part-time                   & -                           & 1                                & 1                    & 1                  \\
P2 & Bachelor                & IN               & Unemployed                  & Blockchain Analyst          & 1                                & 3                    & 0.3                \\
P3 & Graduate School         & PK               & Part-time                   & Freelancer                  & 3                                & 1                    & 1                  \\
P4 & Bachelor                & KE               & Part-time                   & Software Engineer           & 3                                & 2                    & 2                  \\
P5 & Bachelor                & PK               & Part-time                   & Senior Software Developer   & 5                                & 2                    & 2                  \\
P6 & Master                  & IN               & Full-time                   & Blockchain Developer        & 3                                & 1.5                  & 2                  \\
P7 & Vocational degree       & ZW               & Freelancer                  & Developer                   & 5                                & 1                    & 1                  \\ \bottomrule
\end{tabular}}
\end{table*}


\subsection{Interview Guide}
\label{sec:appendix_interview_guide}
{\small
We include only the English version which is identical in content with the German version. 

{\small
\paragraph{Participant onboarding}
\begin{enumerate}
    \item Can you give us a short overview of your role and involvement in the development of Solana smart contracts?
    
    \item What type of applications do you implement? (tokenization, NFTs, ...)
Process of developing Solana smart contract

\end{enumerate}

\paragraph{Process of developing Solana smart contract}
\begin{enumerate}
    \setcounter{enumi}{3}
    \item Can you walk me through the process of creating and developing Solana smart contracts from gathering requirements to implementing and then releasing.
    
    \item During the process of creating and developing Solana smart contracts, is there anything special done to ensure the development of a secure Solana smart contract? (Can you please elaborate.)

    \item What is in your opinion the difference between developing Solana smart contracts and any other software?
 \end{enumerate}
\paragraph{Vulnerabilities in existing Solana smart contracts}  
\begin{enumerate}
\setcounter{enumi}{7}
\item In our analysis of the existing Solana smart contracts, we detected many security vulnerabilities in the deployed Solana smart contracts. Have you encountered a security vulnerability during the development of Solana smart contracts? Can you share with us this experience?

\item Missing Signer check: Can you explain in your own words the vulnerabilities you mentioned and their mitigations.

\item Integer overflow and underflow: Can you explain in your own words the vulnerabilities you mentioned and their mitigations.

\item (In debug the check is enabled / in release mode it is not by default - you would have to enable the checks in Cargo.toml)

\item Arbitrary signed program invocation: Can you explain in your own words the vulnerabilities you mentioned and their mitigations.

\item We spotted a lot of different security vulnerabilities in the deployed Solana smart contracts, in your opinion, what do you think are the reasons that these vulnerabilities are prominent in many Solana smart contracts?

\item In your opinion, which of the existing security vulnerabilities are challenging to find and solve?

\item In your opinion, which of the existing security vulnerabilities are easy to find and solve?

\item In your opinion, what is the most important security vulnerability to look out for?
 \end{enumerate}
\paragraph{Safety of Rust}
\begin{enumerate}
    \setcounter{enumi}{17}
     \item In your opinion, what advantages does Rust offer for the Solana blockchain?
    
    \item How does Rust differ from other programming languages?

    \item If security was not mentioned:
    \begin{enumerate}
    \item What is special about Rust when it comes to security compared to other programming languages?
    \item How does this affect the way you handle security concerns while developing?
    \end{enumerate}

\end{enumerate}

\paragraph{Improving the secure development of Solana smart contracts}
\begin{enumerate}
    \setcounter{enumi}{21}
     \item In your opinion, what kind of improvements can be done to ensure the secure development of Solana smart contracts?
    \item In your opinion, how useful are code reviews when it comes to ensuring the development of secure Solana smart contracts?
    \item What advice would you give to other Solana smart contract developers when it comes to securing Solana smart contracts?
\end{enumerate}

}
}

\section{Overview of Solana}
\label{sec:solana-background}

\begin{figure}[tb]
    \centering
    \includegraphics[bb=0 0 204 144, width=.6\linewidth]{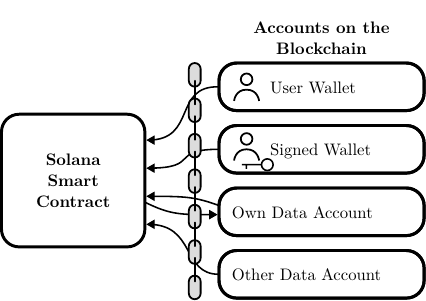}
    \caption{High level overview of Solana.}
    \label{fig:sol_arch}
\end{figure}

Unlike other blockchain platforms like Ethereum, where smart contracts store and update data within the contract itself, Solana's approach to smart contracts follows a stateless design~\cite{solanadoc}. 

In the absence of internal state, Solana incorporates two verification mechanisms during execution to ensure the security of transactions: (i)~signer checks and (ii)~ownership checks.
The signer check is used to verify that the transaction is signed using the key linked to a given user wallet, i.e., the sender possesses the private key associated with the user wallet.
The owner check is used to confirm that a given data account was created and is owned by the executing program.
Since only the owner of a data account can modify the content of its data field, it allows the program to ensure that the provided data account's content can be trusted.

Information on the owner and signer status are provided by the Solana runtime as metadata information of each account.  
However, it is up to the smart contract developer to verify that the checks are used properly for each account to enforce access control.

\section{Neodyme Level 4}
\label{sec:neodyme-4}
Neodyme provided a well-known workshop about Solana smart contract security, in which participants were challenged to develop exploits for a set of vulnerable smart contracts.
The level~4 challenge is a smart contract that contains an \acpi vulnerability.
In general, the vulnerability is similar to the one of the marketplace contract (\Cref{fig:marketplace-cpi}).
In case of the level~4 challenge, the only legitimately intended call target is the SPLToken contract~\cite{spltoken}.
The SPLToken contract is part of the Solana Program Library~\cite{spl} and implements the standard way of handling tokens in the Solana ecosystem.

In its withdraw function, the level~4 challenge uses the SPLToken's token transfer instruction.
However, since the level~4 challenge does not validate the key of its call target, an attacker can redirect the call to \emph{any} smart contract.
For this contract, our symbolic execution tool reports that the call target is the sixth account provided to the smart contract.

\end{document}